\documentclass{agujournal}
\journalname{JGR-Planets}

\usepackage[finalnew]{trackchanges}

\usepackage{nameref}
\usepackage{amsmath}
\usepackage{amssymb}
\usepackage{floatpag}

\usepackage [english]{babel}
\usepackage [autostyle, english = american]{csquotes}
\MakeOuterQuote{"}

\begin{document}

\title{Tidal Evolution of the Evection Resonance/Quasi-Resonance and the Angular Momentum of the Earth-Moon System}

\authors{R. Rufu  and R. M. Canup}
\affiliation{}{Planetary Science Directorate, Southwest Research Institute, Boulder, Colorado, 80302, USA}

\correspondingauthor{R. Rufu}{raluca@boulder.swri.edu}

\begin{keypoints}
\item Evection resonance and a subsequent quasi-resonance regime may remove angular momentum from the Earth-Moon system.
\item Only a narrow range of fixed tidal parameters can reconcile a high-angular momentum Moon-forming impact with the current Earth-Moon system.
\item For a time-dependent terrestrial dissipation model, quasi-resonance escape occurs early, leaving the system with an angular momentum excess.
\end{keypoints}

\begin{abstract} 

Forming the Moon by a high-angular momentum impact may explain the Earth-Moon isotopic similarities, however, the post-impact angular momentum needs to be reduced by a factor \change{$\ge2$}{of 2 or more} to the current value ($1\ L_{\rm EM}$) after the Moon forms. Capture into the evection resonance, occurring when the lunar perigee precession period equals one year, could remove the angular momentum excess. However the appropriate angular momentum removal appears sensitive to the tidal model and chosen tidal parameters. In this work, we use a constant-time delay tidal model to explore the Moon's \add{orbital} evolution through evection. We find that exit from formal evection occurs early and that subsequently, the Moon enters a quasi-resonance regime, in which evection still regulates the lunar eccentricity even though the resonance angle is no longer librating. Although not in resonance proper, during quasi-resonance angular momentum is continuously removed from the Earth-Moon system and transferred to Earth's heliocentric orbit. The final angular momentum, set by the timing of quasi-resonance escape, is a function of the ratio of tidal strength in the Moon and Earth\remove{, $A$,} and the absolute rate of tidal dissipation in the Earth. We consider a physically-motivated model for tidal dissipation in the Earth as the mantle cools from a molten to a partially molten state. We find that as the mantle solidifies, increased terrestrial dissipation drives the Moon out of quasi-resonance. For post-impact systems that contain $>2\ L_{\rm EM}$, final angular momentum values after quasi-resonance escape remain significantly higher than the current Earth-Moon value.
\newline
\textbf{Plain Language Summary}\newline 
Forming the Moon by a high-angular momentum impact may offer a compelling explanation for measured Earth-Moon isotopic similarities. However, the post-impact system after such an event contains a significantly larger angular momentum than the current Earth-Moon system. As the early Moon tidally recedes, its perigee precession rate decreases. When the precession rate equals one year, the Moon may be captured into the evection resonance with the Sun. It was proposed that during this stage the excess angular momentum is removed, but the appropriate angular momentum removal has appeared sensitive to the chosen tidal model. In this work, we find that the Moon exits formal resonance early, but may enter a prolonged quasi-resonance regime, in which angular momentum is continuously removed. The final angular momentum, set by the timing of the quasi-resonance escape, is a function of the relative tidal strength in the Moon and Earth, and the absolute tidal dissipation in the planet. We explore the tidal evolution through evection resonance during planetary cooling adopting a recent model for Earth's time-dependent tidal dissipation, and find that the Moon is driven out of quasi-resonance before sufficient angular momentum is removed, inconsistent with the Earth-Moon value.
\end{abstract}


\section{Introduction}
The leading theory of lunar origin posits that a Mars-sized protoplanet impacted the proto-Earth at the late stages of its accretion \citep{cameron1976origin}. An oblique impact with an angular momentum (AM) close to that in the current Earth-Moon system ($L_{\rm EM}=3.5\cdot 10^{41}\ \rm{g\  cm^2\ s^{-1}}$) generates a debris disk up to two times more massive than the Moon \citep{canup2001origin,canup2004simulations}, which later may accrete to form a lunar-sized  satellite \citep{ida1997lunar,salmon2012lunar}. Such a "canonical" giant impact is able to account for the AM in the current Earth-Moon system and the Moon's depletion in iron and volatile elements \citep{canup2015lunar}.  However, the disk created by a canonical impact is mainly derived from the impactor, while the Earth largely retains its pre-impact composition, which would nominally produce a Moon that is compositionally distinct from Earth's mantle. This is in contrast to high-precision measurements of lunar isotopes, which indicate that the Moon and Earth have essentially identical isotopic compositions in most elements (e.g., in oxygen, \citealp{herwartz2014identification}; titanium, \citealp{zhang2012proto}).

Instead, certain types of  high-AM impacts can create a debris disk and a planet that have nearly equal proportions of impactor material, offering a compelling mechanism to create a satellite that is compositionally similar to the silicate Earth [\citealp{cuk2012making,Canup23112012}; see also \citealp{lock2018origin}]. However, the post-impact Earth-Moon AM is typically $\gtrsim2L_{\rm EM}$ and must be greatly reduced after the Moon forms for such models to be viable.

A promising mechanism to reduce the Earth-Moon system AM to its current value involves a solar resonance called evection \citep{brouwer2013methods,kaula1976lunar,touma1998resonances,cuk2012making}, which occurs when the period of precession of the lunar perigee equals one year. During evection, the angle between the Sun and Moon at lunar perigee (or apogee) maintains a nearly constant value $\sim \pi/2$. As the Moon's orbit expands due to tidal interaction with the Earth, the net solar torque increases the lunar eccentricity and AM is transferred from the Earth-Moon pair to Earth's orbit around the Sun. For an initial 5-hr terrestrial day (corresponding to Earth's spin after a canonical giant impact, \citealt{canup2004simulations}), evection is encountered at 4.6 Earth radii ($R_\oplus$), and only limited AM removal was found \citep{touma1998resonances}. However for a high-AM giant impact (total AM $\gtrsim2\,L_{\rm EM}$), the resonance location shifts outward due to Earth's increased spin and oblateness, and large-scale AM removal was found \citep{cuk2012making}. Notably, for these cases there also appeared to be a preference for a final AM near $\sim 1\,L_{\rm EM}$, independent of the starting AM \citep{cuk2012making}. 

\citet{cuk2012making} used a simplified approximation of a constant lag angle (constant-$Q$) tidal model. Later studies with a full constant-$Q$ model found that the formal evection resonance is unsuccessful at appropriately reducing the Earth-Moon AM \citep{Wisdom2015138,tian2017coupled}. For cases with a large terrestrial tidal dissipation factor, $Q$, it was found that capture into proper evection resonance did not occur, but that instead the Moon was captured into a limit cycle associated with evection, in which appropriate AM can be lost even though the evection resonance angle is not librating \citep{Wisdom2015138,tian2017coupled}.  

Motivated by these differences, we seek to understand whether evection can remove sufficient AM, as this is key for assessing the overall likelihood of high-AM lunar origin scenarios. Recent evection studies \citep{cuk2012making,Wisdom2015138,tian2017coupled} have used N-body codes. In this work we use a complementary semi-analytical model [\citealp{ward2020analytical}; see also \citealp{Ward2013evection}] and test a broader range of tidal parameters, which would be computationally prohibitive with N-body methods.  

In \cite{ward2020analytical}, we developed a simplified model for evection, in which the libration of the resonance angle about $\sim\pi/2$ was assumed to be small. That study estimated that shortly after the Moon's eccentricity becomes large enough to cause its orbit to contract, the libration amplitude increases, and escape from resonance occurs before much AM is removed from the system. However, the approach in \cite{ward2020analytical} could not treat non-librating behavior. In this study, we track the libration of the resonance angle during the encounter with evection resonance, assess the timing of resonance escape and the subsequent evolution, and estimate the final AM  of the system as a function of tidal parameters. We consider both time-constant tidal parameters, and cases in which the terrestrial tidal dissipation factor varies with time as the Earth's mantle cools and begins to solidify \citep{zahnle2015tethered}. 

\section{Model}
We here briefly describe our tidal and evection model (see \citealp{ward2020analytical}, for more details). We assume the Moon forms on a low-eccentricity ($e$) orbit, at a semimajor axis ($a$) slightly beyond the Roche limit\add{, consistent with the lunar orbit after accretion} \citep{ida1997lunar,salmon2012lunar,Salmon20130256}.  The formation of the Moon from a debris disk implies an initially low-inclination, near equatorial orbit \citep{ida1997lunar}, which is inconsistent with the $\sim5^\circ$ inclination of the current Moon's orbit relative to the ecliptic  \citep{goldreich1966history,touma1994evolution,Cuk2016}. Later collisionless encounters of the Moon with the leftover planetesimal population could potentially excite the lunar inclination when it is more distant \citep{Pahlevan:2015aa}. As evection occurs early and close to the planet, we  assume that the lunar inclination relative to the planetary equatorial plane is small throughout the modeled encounter with evection.

We consider the co-planar problem in which the orbital and equatorial planes of the Earth and Moon are aligned. The scalar AM of the Earth-Moon is:
\begin{linenomath*}
\begin{equation}
    L\sim Cs+C_{m}s_{m}+m\sqrt{GM_{\oplus}a(1-e^2)}
    \label{eq:L}
\end{equation}
\end{linenomath*}
where $G$ is the gravitational constant, $M_{\oplus}=5.97\cdot10^{27}\ \rm {g}$ [$m=7.34\cdot10^{25}\ \rm {g}$], $C$ [$C_m$], and $s$ [$s_m$] are Earth's [Moon's] mass, moment of inertia, and spin rate, respectively (here we assume that $(1+m/M_{\oplus})^{-1/2}=0.994\sim1$). We normalize the AM by $C\Omega_\oplus$, where $\Omega_\oplus=\sqrt{GM_\oplus/R^3_\oplus}$ is the orbital frequency at 1 Earth radius ($R_\oplus$), giving
\begin{linenomath*}
\begin{equation}
    L^{\prime}= s^{\prime}+\kappa s^{\prime}_m+\gamma\sqrt{a^{\prime}(1-e^2)},
    \label{eq:Lprime}
\end{equation}
\end{linenomath*}
where primes indicate normalized values ($s^{\prime}=s/\Omega_\oplus$, $s^{\prime}_m=s_m/\Omega_\oplus$, and $a^{\prime}=a/R_\oplus$), $\kappa\equiv C_m/C=1.07\cdot10^{-3}$ is the ratio of the maximum principal moments of inertia of the two bodies, 
$\gamma\equiv\mu/\lambda=0.0367$, with $\mu$ being the Moon-to-Earth mass ratio and $\lambda\equiv C/M_{\oplus}R_{\oplus}^2=0.335$ the Earth's gyration constant. Specifically, $L^{\prime}$ is the ratio of the total AM to that of a single object having Earth's mass and moment of inertia, rotating at $\Omega_{\oplus}$, which is close to the critical rotation rate before break-up. In these units, the current Earth-Moon AM is $L^{\prime}_{\rm EM}\approx0.35$.

Tidal interactions between the Earth and Moon exchange AM between the objects' spins and the lunar orbit, while maintaining the total AM constant. Tides raised on the Earth by the Moon (Earth tides) alter $s^{\prime}$, while tides raised on the Moon by Earth  (lunar tides) alter $s^{\prime}_m$ (we nominally assume a non-synchronous lunar rotation; see section 3.3 for cases assuming synchronous lunar rotation). The tidal changes to Earth's ($s^{\prime}$) and Moon's ($s^{\prime}_m$) spin are:
\begin{linenomath*}
\begin{equation}
    \dot{s}^{\prime}=-\frac{1}{2}\gamma\sqrt{a^{\prime}(1-e^2)}\left( \frac{\dot{a}^{\prime}_\oplus}{a^{\prime}} - \frac{2e\dot{e}_\oplus}{1-e^2}
    \right)
\label{eq:s}
\end{equation}
\begin{equation}
    \dot{s}^{\prime}_m=-\frac{1}{2}\frac{\gamma}{\kappa}\sqrt{a^{\prime}(1-e^2)}\left(\frac{\dot{a}^{\prime}_m}{a^{\prime}} - \frac{2e\dot{e}_m}{1-e^2}\right)
    \label{eq:sm}
\end{equation}
\end{linenomath*}
with the $\dot{a}^{\prime}_\oplus$ and $\dot{e}^{\prime}_\oplus$ [$\dot{a}^{\prime}_m$ and $\dot{e}^{\prime}_m$] representing rates of change due to Earth [lunar] tides. 

The time derivatives in eqn. (\ref{eq:s}) and (\ref{eq:sm}) (and henceforth) use a normalized time variable $\tau=t/t_T$, where $t_T$ is a tidal time constant and, defined by $t_T\equiv(6k_{2\oplus}\mu \Omega^2_{\oplus}\Delta t)^{-1}$, where $\Delta t$ is the terrestrial tidal time delay (see below), and $k_{2\oplus}$ is Earth's Love number, where we set \change{$k_{2\oplus}=0.25$}{$k_{2\oplus}\sim0.3$} \citep{murray1999solar}. For example, the current terrestrial tidal factor $Q\sim12$ \citep{murray1999solar} corresponds to a time lag of $\Delta t\sim 10\ \rm {min}$ and \change{$t_T\sim15\ \rm{hr}$}{$t_T\sim13.6\ \rm{hr}$} for the current $s$ and $n$ of the Earth-Moon \add{(where $n$ is the lunar mean motion)}.

All common tidal models used for long-term integrations make simplifying assumptions. We employ the tidal model developed by \cite{mignard1979evolution,mignard1980evolution}, which  makes two main assumptions.  First, that the tidal distortion due to the perturbing body can be accurately described as an additional second-order term ($l=2$) in the potential of the perturbed body.  Second, that in each body, the formation of the equilibrium tide is delayed by some time, $\Delta t$, relative to the tide-raising perturbation, and that this time delay does not depend on the frequency of the tidal response. The time delay reflects  the  effects  of  dissipation  in  the  distorted  body, which can be different for each body (i.e., $\Delta t$ for the Earth can be different than $\Delta t_m$ for the Moon). A key advantage of the Mignard model is its physically intuitive treatment for eccentric orbits and those near the co-rotation distance (where the Moon's orbital period equals the Earth's day), both conditions that can apply to evolution in evection (see Text S1 for more details). In the Mignard model, the orbit-averaged tidal lag angle, $\delta$, between the equilibrium Earth tide and the sub-lunar position is given by $\delta=(s-n)\Delta t$\remove{ (where $n$ is the lunar mean motion)}. The lag angle is then frequency-dependent, and the resulting torque smoothly approaches zero for a circular orbit as co-rotation is approached and $n\rightarrow s$. In contrast, a constant-$Q$ model (e.g., \citealp{kaula1964tidal}) assumes the tidal response is delayed by a fixed phase, rather than by a fixed time, relative to the tide-raising potential. A difficulty is that the resulting lag angle varies discontinuously as the frequency associated with each term passes through zero; e.g., for the semi-diurnal tide on the Earth, the lag angle abruptly changes from a positive (leading) to a negative (trailing) value for orbits just outside to just inside the co-rotation radius, and there is no variation in the size of the lag angle based on the closeness of the orbit to co-rotation (e.g., \citealp{Cuk2016}). 

The known responses of the current Earth and Moon can be compared to predictions of tidal models. For a tidal dissipation factor, $Q$, that varies with frequency $\chi$ as $Q\propto \chi^{\tilde{\alpha}}$, the Mignard model predicts $\tilde{\alpha}=-1$, while the constant-$Q$ model predicts $\tilde{\alpha}=0$. Seismic measurements imply $\tilde{\alpha}= 0.2$ to $0.4$ for the current terrestrial mantle (e.g., \citealp{efroimsky2007physics}), while analyses of lunar laser ranging (LLR) data suggest $\tilde{\alpha}$ values that are small but negative for the current Moon \citep{williams2014lunar,williams2015tides}. Thus LLR data seems better described by a constant-$Q$ model, although tidal models in which dissipation peaks at a certain  frequency (plausibly simulating the effect of a low-viscosity layer) seem to most successfully fit the current lunar measurements \citep{williams2015tides}. However, the extent to which these results bear on the initial Earth and Moon is unknown, particularly for the post-giant impact Earth that was likely molten and fluid-like in its tidal response (see below). We adopt the Mignard model as a reasonable proxy for the initial Earth-Moon system. The Mignard lag time can be related to a tidal quality factor, $Q$, as $Q \sim (\psi \Delta t)^{-1}$, where $\psi$ is the frequency of oscillation.  For the Earth, the \change{dominant}{semidiurnal lunar tide} frequency is $2|s-n|$, where the factor of 2 arises because there are two tidal cycles for each synodic period (e.g., \citealp{efroimsky2007physics,peale2015origin}). For ease of comparison with previous works, we express results as a function of an effective terrestrial tidal dissipation factor, $Q_{\rm eff}$, where $Q_{\rm eff}=(2(s_0-n_0)\Delta t)^{-1}$,  and $s_0$ and $n_0$ are the initial values of $s$ and $n$ in our simulations. However, our actual simulations utilize $\Delta t$ and $A$ as defined in eqn. \ref{eq:A} below, rather than $Q_{\rm eff}$.

\citeauthor{mignard1980evolution}'s (\citeyear{mignard1980evolution}) normalized equations for the rates of change of $a^{\prime}$ and $e$ due to Earth's tidal dissipation (Earth tides) read (e.g., \citealp{ward2020analytical}):
\begin{linenomath*}
\begin{equation}
    \frac{\dot{a}^{\prime}_\oplus}{a^{\prime}}=(\mu+1)\left( \frac{s^{\prime}a^{\prime 3/2}f_1(e)-f_2(e)}{a^{\prime 8}} \right)
    \label{eq:aplus}                                           
\end{equation}
\begin{equation}
    \dot{e}_\oplus=e(\mu+1)\left( \frac{s^{\prime}a^{\prime 3/2}g_1(e)-g_2(e)}{2a^{\prime 8}} \right)
    \label{eq:eoplus}
\end{equation}
\end{linenomath*}
where $f_1$, $f_2$, $g_1$, and $g_2$ are functions of $e$ found by averaging the tidal forces over one lunar orbit (see Table S1; \citealp{mignard1980evolution,meyer2010coupled,ward2020analytical}).

For non-synchronous lunar rotation, the corresponding rates due to lunar tidal dissipation (lunar tides) are:
\begin{linenomath*}
\begin{equation}
    \frac{\dot{a}^{\prime}_m}{a^{\prime}}=A(\mu+1)\left( \frac{s^{\prime}_m a^{\prime 3/2}f_1(e)-f_2(e)}{a^{\prime 8}} \right)
    \label{eq:am}
\end{equation}
\begin{equation}
    \dot{e}_m=eA(\mu+1)\left( \frac{s^{\prime}_m a^{\prime 3/2}g_1(e)-g_2(e)}{2a^{\prime 8}} \right)
    \label{eq:em}
\end{equation}
\end{linenomath*}
where $A$ is the relative strength of lunar tides compared to Earth tides, defined as:
\begin{linenomath*}
\begin{equation}
    A\equiv \frac{k_{2m}}{k_{2\oplus}}\frac{\Delta t_m}{\Delta t}\left(\frac{M_\oplus}{m}\right)^2\left (\frac{R_m}{R_\oplus}\right)^5,
    \label{eq:A}
\end{equation}
\end{linenomath*}
where $k_{2\oplus}$ [$k_{2m}$], $\Delta t$ [$\Delta t_m$], and $R_\oplus$ [$R_m$] are Earth's [Moon's] Love number, tidal lag time, and radius respectively. When $A\gg1$, tidal dissipation in the Moon is relatively stronger than in the planet, and hence in this case, lunar tides (which typically decrease $a$ and $e$) are stronger than Earth tides (which typically increase $a$ and $e$). Shortly after the giant impact, $A$ was likely $\gg1$, because Earth was fully molten and surrounded by a thick atmosphere (leading to a fluid-like response with a small $\Delta t$), whereas the Moon would have cooled more quickly to a dissipative partially solid state (with a larger $\Delta t_m$) by the time the Moon encountered the evection resonance (\citealp{zahnle2015tethered}; see section 3.4). 

\begin{figure}
    \centering
    \includegraphics[width=0.5\linewidth]{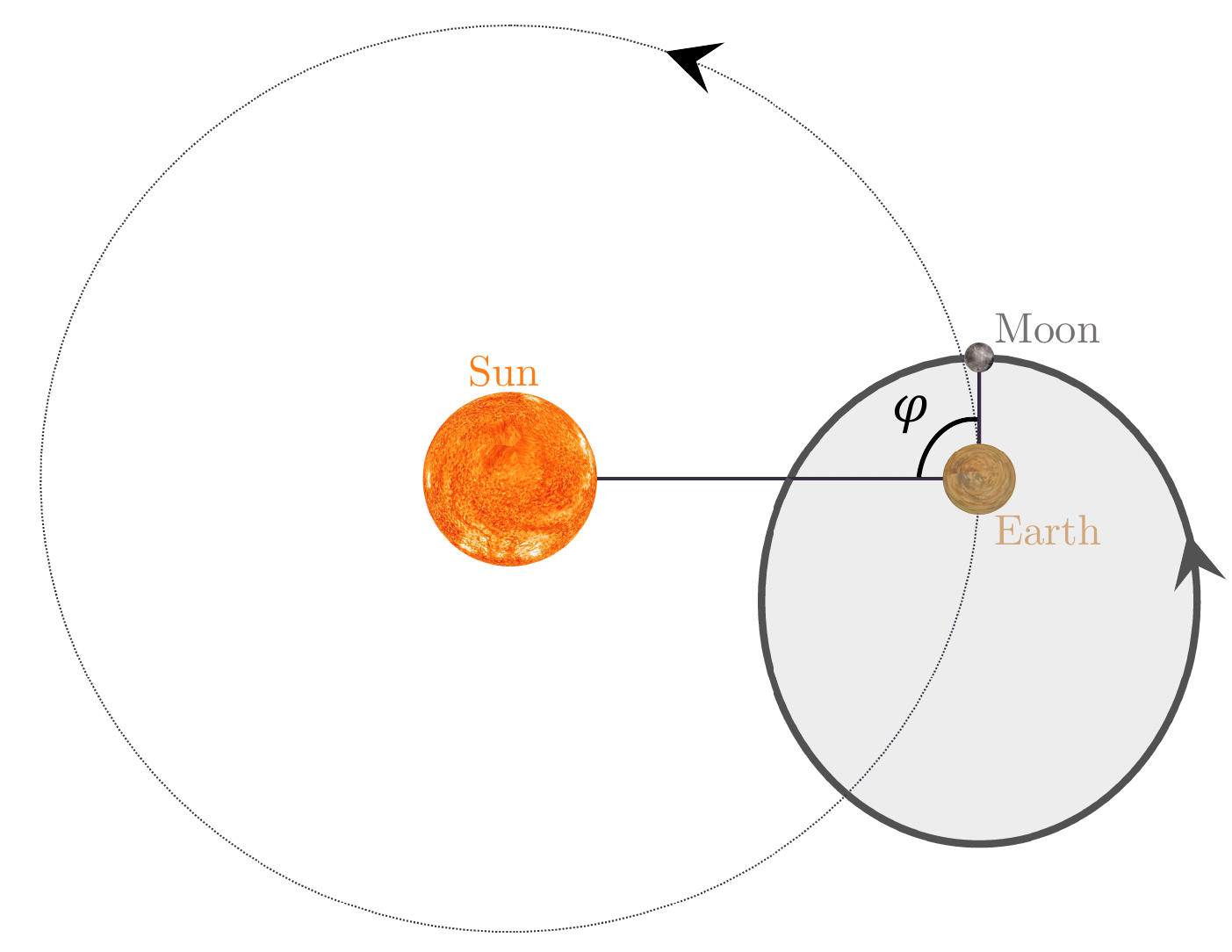}
    \caption{\justifying Schematic of the Earth's (dashed line) and lunar (solid line) orbit. During proper evection, the angle between the Sun and the Moon at perigee (with Earth as the reference point) librates about $\varphi\sim\pm\pi/2$, with perigee occurring either on the leading side compared to Earth's heliocentric motion (as seen here) or on the trailing side. Distances and sizes are not to scale.}
    \label{fig:Schematic}
\end{figure}

The Moon initially evolves outward due to tides until it reaches the semimajor axis at which resonance occurs ($a^{\prime}_{\rm res}$). The effect of evection on the lunar eccentricity is \citep{brouwer2013methods,ward2020analytical}:
\begin{linenomath*}
\begin{equation}
    \frac{de_{\rm res}}{d\tau}=\dot{e}_{\rm res} = \frac{15}{4}e\sqrt{(1-e^2)}a^{\prime3/2}(\Omega_\odot t_T)\frac{\Omega_\odot}{\Omega_\oplus}\sin(2\varphi),
    \label{eq:eres}
\end{equation}
\end{linenomath*}
where $\varphi=\varpi-\lambda_\odot$ is the angle between the Moon's longitude of perigee, $\varpi$, and the solar longitude, $\lambda_\odot$, with Earth as the reference point (Fig. \ref{fig:Schematic}), and $\Omega_\odot=\dot{\lambda}_\odot$ is Earth's orbital frequency.

The phase angle, $\varphi$, is altered by solar interactions and Earth's quadrupole gravitational field \citep{ward2020analytical}:
\begin{linenomath*}
\begin{equation}
    \frac{d\varphi}{d\tau}=\dot{\varphi} =  \left[\frac{\Lambda^2s^{\prime2}}{a^{\prime7/2}(1-e^2)^2}-1+\frac{3}{4}\sqrt{1-e^2}a^{\prime3/2}\left(\frac{\Omega_\odot}{\Omega_\oplus} \right) (1+5\cos{2\varphi}) \right]\Omega_\odot t_T,
    \label{eq:phi}
\end{equation}
\end{linenomath*}
where we assume that Earth's oblateness is a function of its spin, $J_2=J_\star s^{\prime 2}$ ($J_\star=0.315$), and define $\Lambda\equiv\sqrt{3/2J_\star\Omega_\oplus/\Omega_\odot}=54.2$. 
In the vicinity of evection, when $a^{\prime}_{\rm res}=\left [\Lambda s^{\prime}/(1-e^2)\right ]^{4/7}$, the apsidal precession rate, $\dot{\varpi}$, approaches Earth's orbital frequency, $\Omega_\odot$, and the first term on the right-hand side in eqn. (\ref{eq:phi}) is $\sim1$. In that case, the resonance angle librates slowly because the term proportional to $\Omega_\odot/\Omega_\oplus$ is $\ll1$.

\begin{figure}
     \centering
     \includegraphics[width=1.00\linewidth]{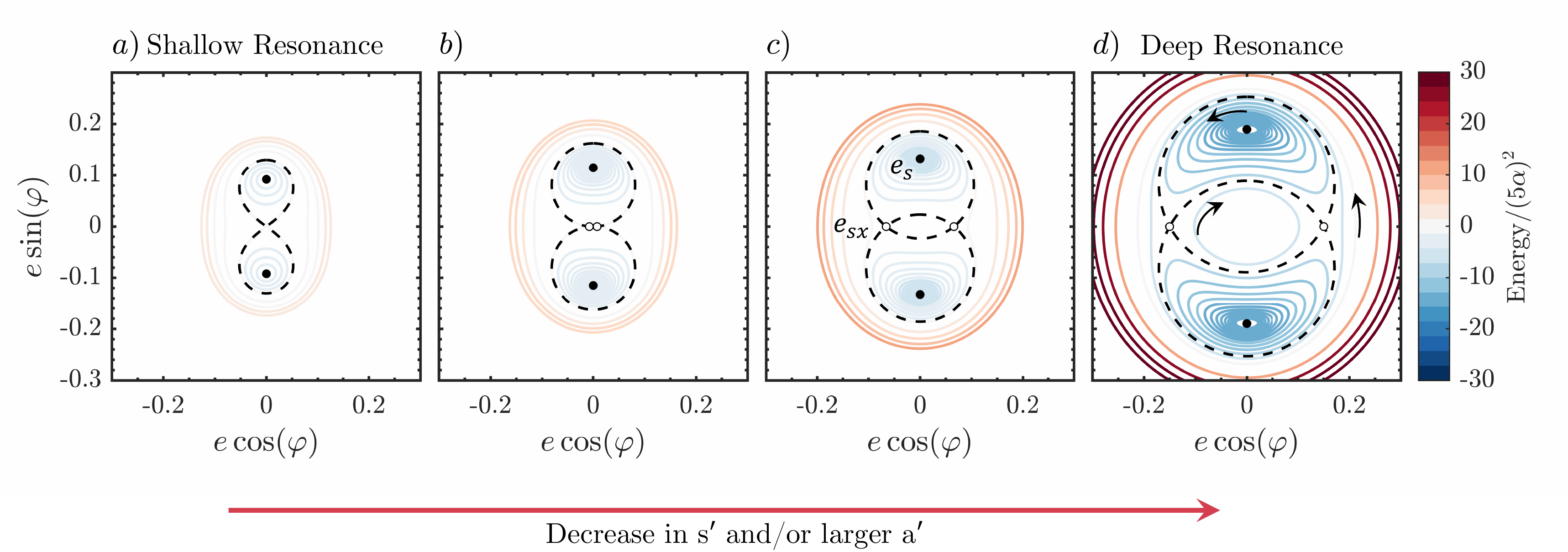}
     \caption{\justifying Example level curves for different energies (depicted as different colors). The Sun is in the direction of the positive $x$-axis. The small full [empty] circles represent the stable [unstable] stationary points, $e_s$ [$e_{sx}$]. The dashed curves are the separatrix that divides the phase space into as many as three regions: libration around one stable stationary point (evection proper), circulation around the origin, and circulation around all stationary points. The panels show different evolution stages: a) shallow resonance, when the unstable stationary point is not defined (eqn. \ref{eq:esx} has no real solution); b) As the planetary spin, $s^{\prime}$, decreases and/or the lunar semimajor axis, $a^{\prime}$ increases, the resonance  further develops ($e_s$ and $e_{sx}$ increase). c) Deep resonance, defined as the stage when $e_{sx}>0$ and the central circulating region between the unstable saddle points emerges. d) Further evolution causes the central area between the saddle points to occupy an increasing  portion of the phase diagram. Arrows in panel d) show the direction of trajectory motion. The energy levels are normalized by $(5\alpha)^2$, where $\alpha\equiv3/8a^{\prime3/2}\Omega_\odot/\Omega_\oplus$.}
    \label{fig:LevelCurves}
 \end{figure}

For given values of $s^{\prime}$ and $a^{\prime}$, one can find the stationary points at which $\dot{e}_{\rm res}$ and $\dot{\varphi}$ vanish. From eqn. (\ref{eq:eres}) we must have $\varphi\approx0,\pm\pi/2,\pi$, and from eqn. (\ref{eq:phi}) the stable stationary ($e_s$) and unstable saddle ($e_{sx}$) eccentricity points solve the equations:
\begin{linenomath*}
\begin{equation}
    \frac{\Lambda^2s^{\prime2}}{a^{\prime7/2}(1-e_s^2)^2}-1-3\sqrt{1-e_s^2}a^{\prime3/2}\frac{\Omega_\odot}{\Omega_\oplus}=0
    \label{eq:es}
\end{equation}
\begin{equation}
    \frac{\Lambda^2s^{\prime2}}{a^{\prime7/2}(1-e_{sx}^2)^2}-1+\frac{9}{2}\sqrt{1-e_{sx}^2}a^{\prime3/2}\frac{\Omega_\odot}{\Omega_\oplus}=0
    \label{eq:esx}
\end{equation}
\end{linenomath*}
Finally $e=0$ is also a stationary point for which $\dot{e}_{\rm res}=0$ and $\varphi$ is undefined. In the absence of tidal dissipation, the solar terms do not alter the overall energy of the system, hence an integral of motion defines the allowed values of $(e,\varphi)$. Level curve diagrams for different energies are shown in Fig. \ref{fig:LevelCurves}.


Initially after the Moon encounters evection, $e_s$ is small and $e_{sx}$ is not defined (i.e., eqn. \ref{eq:esx} has only imaginary solutions), corresponding to what we will call shallow resonance (Fig. \ref{fig:LevelCurves}-a). As Earth's spin, $s^{\prime}$, decreases and/or $a^{\prime}$ increases,  $e_s$ increases as well (moves up the $y$-axis) and an unstable point appears at the origin ($e_{sx}=0$; Fig. \ref{fig:LevelCurves}-b). In shallow resonance, there are two regimes: for low energy levels (blue colors in Fig. \ref{fig:LevelCurves}-b), trajectories librate around a stable point $(e_s,\pm\pi/2)$, while higher energies (red colors in Fig. \ref{fig:LevelCurves}-b) circulate around both stable points. Further in the evolution, in what we refer to as deep resonance, at a certain energy level the trajectory intersects the saddle points $(e_{sx},0),(e_{sx},\pi)$, dividing the phase space into three regions (separatrix curve; dashed line in Fig. \ref{fig:LevelCurves}-c). As continued evolution drives the level curves outward, the inner region around the origin occupies a larger part of the phase space (Fig. \ref{fig:LevelCurves}-d). In this inner region, trajectories circulate about the origin in a clockwise motion, while beyond the outer separatrix, trajectories circulate both stationary points in a counter-clockwise motion. The intermediate, increasingly crescent-like region is that of resonant libration, in which trajectories travel in an overall counter-clockwise sense.

As seen from the snapshots in Fig. \ref{fig:LevelCurves}, the trajectories are symmetric around the $y$-axis, therefore, in the absence of tidal evolution, evection alone does not alter the AM of the system because there is no net solar torque ($T\propto e^2\sin{2\varphi}$; \citealt{ward2020analytical}). However, the gradual evolution of the level curves, due to tidal dissipation, disrupts the $y$-axis symmetry (i.e., the actual trajectories are no longer closed), and AM is removed from the system. While \textit{not} in evection resonance, Earth and lunar tides alter the semimajor axis and eccentricity (eqns. \ref{eq:aplus}-\ref{eq:em}), which is directly balanced by the changes in Earth's and lunar spins (eqns. \ref{eq:s} and \ref{eq:sm}), so that the total AM of the system (eqn. \ref{eq:L}) remains constant. During evection resonance, tides still control the semimajor axis and spin evolutions, but so long as the rate of change of the eccentricity is controlled by evection (eqn. \ref{eq:eres}), the total AM is no longer constant. Instead, AM from the Earth-Moon system is transferred to Earth's orbit around the Sun.

To model the Earth-Moon system as it tidally evolves and encounters evection, we integrate the equations for the Earth and lunar spin rates (eqns. \ref{eq:s} and \ref{eq:sm}), the Moon's semimajor axis ($\dot{a}=\dot{a}_\oplus+\dot{a}_m$, using eqns. \ref{eq:aplus} and \ref{eq:am}), eccentricity ($\dot{e}=\dot{e}_\oplus+\dot{e}_m+\dot{e}_{\rm res}$, using eqns. \ref{eq:eoplus}, \ref{eq:em} and \ref{eq:eres}) and the resonance angle (eqn. \ref{eq:phi}). We use the "VODE" (Variable-coefficient Ordinary Differential Equation, \citealp{brown1989vode}) integrator in \textit{SciPy} package  \citep{virtanen2019scipy}. 
We assume a minimum eccentricity of $e=10^{-8}$, which approximates the minimum eccentricity expected due to excitation from collisionless encounters when the Moon's semimajor axis is small (e.g. \citealp{spurzem2009dynamics, Pahlevan:2015aa}). The derivative at the minimum eccentricity is set to $\dot{e}=\dot{e}_{res}$, limiting further eccentricity decrease due to tides but allowing the solar terms to increase the eccentricity when evection occurs 
(see Text S2 in Supplementary material). We stop the integration if the Moon's perigee is $<2R_{\oplus}$ (a tidal disruption boundary for high-$e$ orbits, \citealp{sridhar1992tidal}) or if the lunar spin rate, the Earth's spin rate, and the Moon's mean motion are equal (i.e., if the system reaches the double synchronous state).

\section{Results}
At the start of each simulation, $e=0.01$ (e.g., \citealp{salmon2012lunar}) and $a^\prime=3.5$. We assume an initial terrestrial spin of $2\ \rm hr$, which corresponds to a total AM of $L^{\prime}=0.77$ (i.e., $L=2.2\,L_{\rm EM}$, consistent with high-AM Moon forming impacts, \citealp{cuk2012making,Canup23112012}). The tidal time constant, $t_T$ (which is proportional to $\Delta t^{-1}\propto Q_{\rm eff}$), the relative tidal strength, $A$, and $0\leq\varphi(0)\leq2\pi$ are input parameters.

The remainder of section 3 is organized as follows. First, we describe in section 3.1 the evolution  assuming that the Moon's rotation evolves solely due to tides (implying non-synchronous lunar spin for eccentric orbits) and for relatively low $A$ and $\Delta t$ values that are constant in time, similar to conditions considered in \cite{cuk2012making} (their Figure 4-b; purple line). Next, we compare the final AM after encounter with evection across a large range of tidal parameters, again assuming that $A$ and $\Delta t$ are constant in time throughout the evolution (for non-synchronous lunar rotation in section 3.2, and for synchronous lunar rotation in section 3.3). In section 3.4, we consider the most physically plausible initial condition with large $A$ and large $Q_{\rm eff}/k_{2\oplus}$, and we adopt a time-dependent terrestrial dissipation model to describe the Moon's evolution through evection resonance during Earth's gradual solidification after a giant impact \citep{zahnle2015tethered}.

\subsection{Resonance and Quasi-Resonance Evolution}
\begin{figure} 
\thisfloatpagestyle{empty}
     \centering
     \includegraphics[width=0.70\linewidth]{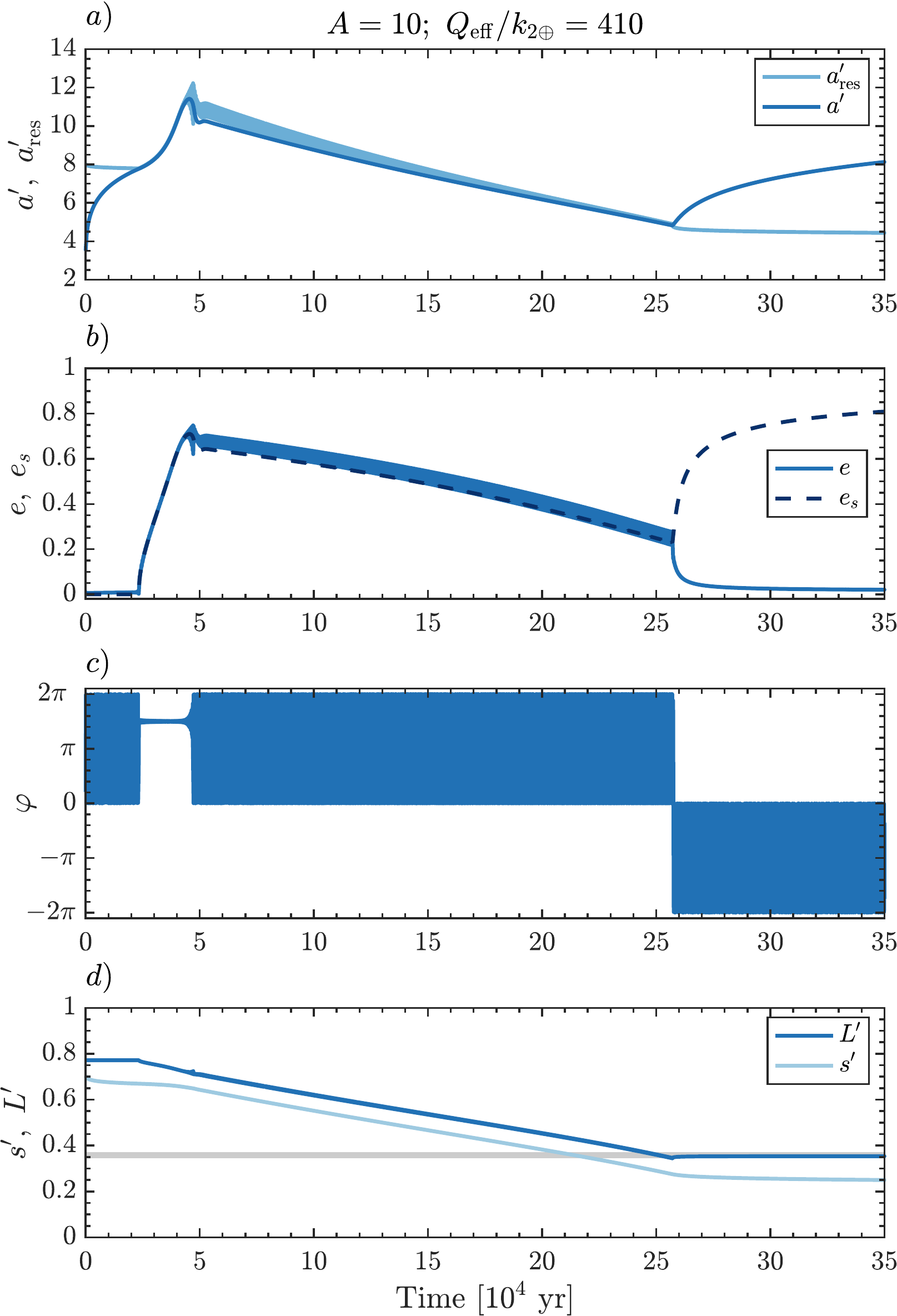}
     \caption{\justifying Tidal evolution of the Earth-Moon system for $A=10$ and terrestrial tidal parameters $Q_{\rm eff}/k_{2\oplus}=410$ ($t_T=4.9\times10^6 \ \rm sec$). We assume that the Moon's rotation evolves solely due to tides (i.e., non-synchronous lunar spin for eccentric orbits). a) Normalized lunar semimajor axis (dark blue) and position of evection resonance (light blue) semimajor-axis; b) eccentricity (solid line) and stable stationary eccentricity (dashed line - numerically calculated by solving eqn. \ref{eq:es}); c) resonance angle d) normalized AM (dark blue) and terrestrial spin (light blue) as a function of time (see text for details). For the initial AM  of $L^\prime=0.77$ ($L\sim2.2L_{\rm EM}$), evection resonance is encountered at $a=7.8\,R_{\oplus}$. The horizontal grey area  represents values consistent with the current Earth-Moon, accounting for later AM change due to solar tides and late accretion impacts  \citep{canup2004simulations,bottke2010stochastic}. We set negative values for the resonance angle, $\varphi$, when the derivative $\dot{\varphi}$ is negative, indicating that the lunar precession period is $>1\,\rm yr$ and that the trajectory moves in the clockwise direction on the phase diagram in the central circulation region. For comparison today's values are \change{$A\sim13$}{$A\sim13$} and $Q_{\rm eff}/k_{2\oplus}\sim40$ \citep{murray1999solar,williams2015tides}}
     \label{fig:EvolutionFastTides}
\end{figure}

Fig. \ref{fig:EvolutionFastTides} shows results from an integration with $A=10$ and $Q_{\rm eff}/k_{2\oplus}\approx410$. Initially, the Moon's orbit expands at a low eccentricity due to Earth tides, until it is captured in the evection resonance at $a^\prime=7.77$ (Fig. \ref{fig:EvolutionFastTides}-a). The resonance drives an increase in eccentricity (solid line in Fig. \ref{fig:EvolutionFastTides}-b) as the orbit further expands due to Earth tides ($|\dot{a}_{\oplus}|>|\dot{a}_m|$). During this outbound phase, $e$ closely follows the stable stationary eccentricity, $e_s$ (dashed line in Fig. \ref{fig:EvolutionFastTides}-b), and the resonance angle librates about $\varphi\sim\pi/2$ with small amplitude (Fig. \ref{fig:EvolutionFastTides}-c). The trajectory is librating within the phase diagram region surrounding the stable stationary point with a counter-clockwise motion (Fig. \ref{fig:LevelCurves}). Only modest AM ($\lesssim10\%$) is removed during this phase.


As the eccentricity grows further, lunar tides become stronger, eventually overcoming the orbital expansion driven by Earth tides and causing the semimajor axis growth to stall when $a^\prime\approx11.4$ and $e\approx 0.7$. From this point, the orbit contracts ($|\dot{a}_{\oplus}|<|\dot{a}_m|$), the libration amplitude increases until it reaches the saddle point $(e_{sx},0)$, and the trajectory exits the resonant region. In this simulation, escape occurs to the high-$e$ side into the region beyond the outer separatrix (Fig. \ref{fig:LevelCurves}), where the resonance angle no longer librates but instead circulates across all values between $0$ and $2\pi$ (Fig. \ref{fig:EvolutionFastTides}-c) in a counter-clockwise sense on the phase diagram. The Moon is now not in evection proper, but it enters a prolonged phase in which evection still regulates the secular evolution of the Moon's eccentricity. We call this state quasi-resonance (QR).

In QR, the Moon's precession period is somewhat less than $1$ yr, and the Moon's orbit is interior to the position of the resonance. The Moon's orbit continues to contract inward due to the predominance of lunar tides. However, as the lunar semimajor axis decreases, the position of evection resonance ($a_{\rm res}$, light blue Fig.\ref{fig:EvolutionFastTides}-a) moves inward at a somewhat faster rate because of the ongoing decrease in Earth's spin, $s^{\prime}$, and the associated reduction in Earth's $J_2$. Hence with time the resonance converges on the Moon's orbit from the outside. During QR, tides drive the trajectory downward on the phase diagram (Fig. \ref{fig:LevelCurves}), but they are blocked by evection when it approaches the separatrix from above, which forces the trajectory to maintain a minimum eccentricity of $\approx e_{sx}$. So long as the tidal change is slow compared to the circulation period, the evolution is adiabatic and the system continues to circulate just above the separatrix. Even though $\varphi$ is circulating, because evection continues to maintain $e\gtrsim e_{sx}$, AM is transferred from the Earth-Moon system to Earth's orbit. It is during this QR regime that substantial AM is removed (Fig. \ref{fig:EvolutionFastTides}-c).

In QR, the trajectory closely hugs the outer separatrix as $\varphi$  circulates from $0$ to $2\pi$, with the eccentricity oscillating between a minimum value $e_{\rm min}\sim e_{sx}$ when near the $x$-axis on the phase diagram, and a maximum value $e_{\rm max} \sim e_+$ comparable to that of the outer separatrix curve near the $y$-axis on the phase diagram (see Fig. S3). To the order $e^4$ and for deep resonance, the later can be expressed as $e_+^2=e_s^2+\sqrt{5\alpha e^2_*}$, where  $e^2_*\equiv(e^2_s+e^2_{sx})/2\sim1-\alpha-\Lambda s^{\prime}/a^{\prime7/4}$, and $\alpha=3/8a^{\prime 3/2}\Omega_\odot/\Omega_\oplus$ \citep{ward2020analytical}. For shallow resonance, the unstable stationary eccentricity is not defined and $e_+^2=2e^2_s$ \citep{ward2020analytical}. At the level of an order $e^4$ approximation, eqn. (\ref{eq:phi}) can be expressed as: $\dot{\varphi} = 2\Omega_\odot t_T\left(e^2-e^2_*+5\alpha\cos{2\varphi}\right)$ \citep{ward2020analytical}, implying an angle-averaged circulation rate $\left<\dot{\varphi}\right>=2\Omega_\odot t_T\left( \left<e^2\right>-e^2_*\right)$, where for deep resonance, $\left<e^2\right>\sim(e^2_{\rm min}+e^2_{\rm max})/2\sim e^2_*+\sqrt{5\alpha e^2_*}$ (or for shallow resonance, $\left<e^2\right>\sim e^2_*+5\alpha$). The time to complete an eccentricity oscillation cycle is:
\begin{linenomath*}
\begin{equation}
 P_{\rm{circ;deep}}=\pi/\dot{\varphi}=\frac{\pi}{2t_T\Omega_{\odot}\sqrt{5\alpha}e_*};\hspace{0.02\linewidth}
 P_{\rm{circ;shallow}}=\frac{\pi}{2t_T\Omega_{\odot}{5\alpha}},
 \label{eq:libP}
\end{equation}
\end{linenomath*}
where $P_{\rm{circ;deep}}$ [$P_{\rm{circ;shallow}}$] refers to the period of oscillations during deep resonance [shallow resonance].

\begin{figure}
    \centering
     \includegraphics[width=1\linewidth]{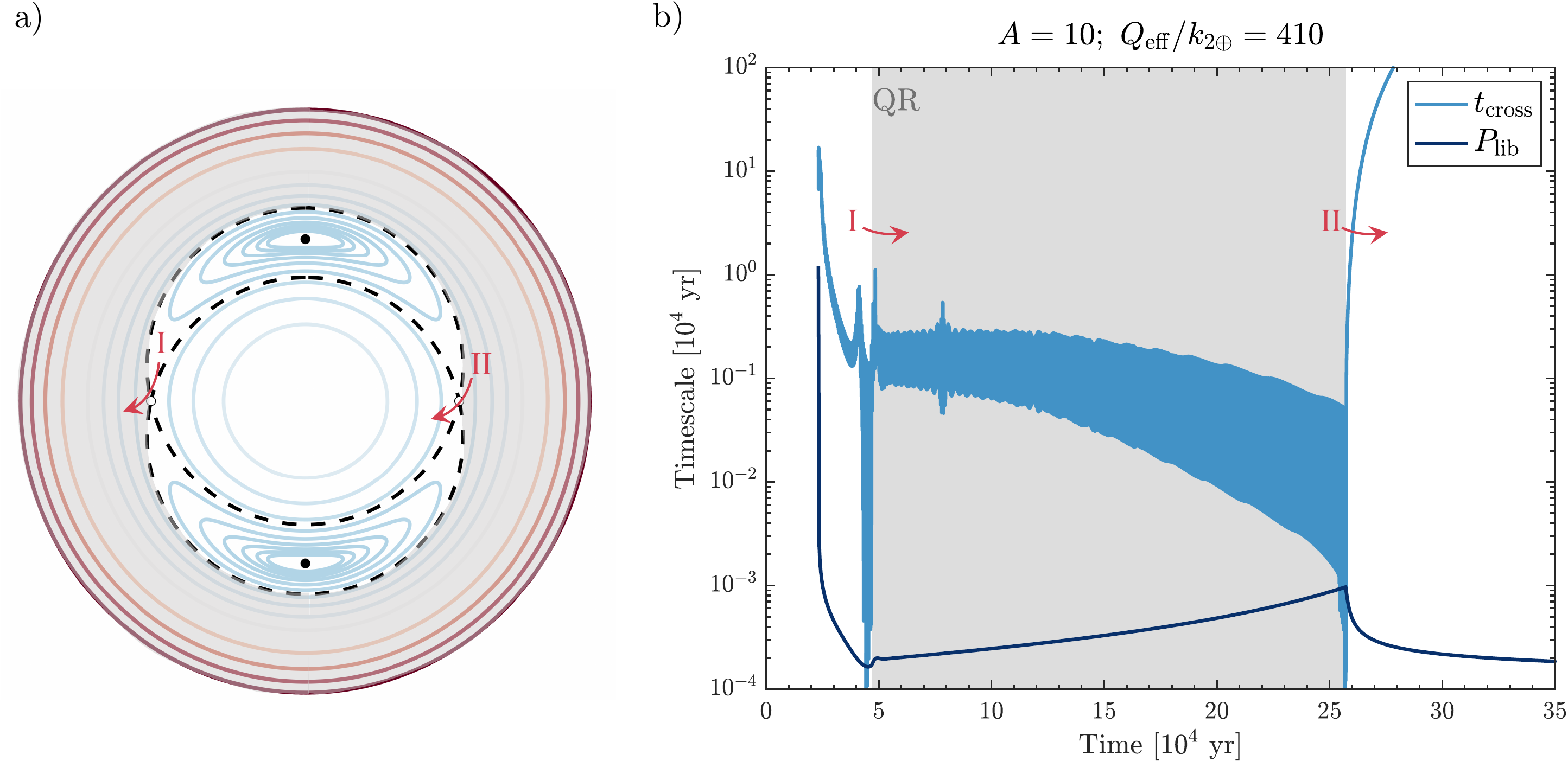}
     \caption{\justifying Evolution from resonance to quasi-resonance. a) Schematic of the level curves (similar to Fig. \ref{fig:LevelCurves} with the QR region highlighted by the grey area. Arrow I [II] depicts the path taken to exit evection proper [QR regime]. b) The timescale required for tides to drive the eccentricity across the separatrix ($t_{\rm cross}$, light blue; calculated using eqn. (\ref{eq:Tcross}) and averaging over a $100$ yr period) compared to the timescale of circulation ($P_{\rm circ}$; dark blue) for the evolution depicted in Fig. \ref{fig:EvolutionFastTides} ($A=10$ and $Q_{\rm eff}/k_{2\oplus}=410$). The grey area represents the time the system is in the QR regime. When the two timescales are comparable, the trajectory is pushed across the separatrix. At $\sim4\cdot10^4\ {\rm yr}$ the libration amplitude increased and  the system is pushed from the librating resonant area across the separatrix, exiting proper evection and entering the QR-regime (arrow I). At $\sim25\cdot10^4\ \rm {yr}$ tides drive the system across the separatix into the central region where clockwise circulation occurs around the origin (arrow II). Once this occurs, the QR-regime ends.}
      \label{fig:CrossLib}
 \end{figure}

The timescale for tides to cause the eccentricity to decrease to a value less than $e_{sx}$ is approximately:
\begin{linenomath*}
\begin{equation}
 t_{\rm cross}=(e-e_{sx})/|\dot{e}_\oplus+\dot{e}_m|
 \label{eq:Tcross}
\end{equation}
\end{linenomath*}
While the circulation timescale (eqn. \ref{eq:libP}; dark blue line Fig. \ref{fig:CrossLib}-b) increases as $e$ and $a^{\prime}$ decrease, the tidal crossing timescale (eqn.  \ref{eq:Tcross}; light blue line Fig. \ref{fig:CrossLib}-b) generally decreases as the eccentricity gradually approaches $e_{sx}$. When these timescales become comparable, tides drive the Moon inward across the separatrix into the inner non-resonant region when $\varphi\approx0$ or $\pi$ (i.e., when the trajectory is near the $x$-axis on the phase diagram).  This is the end of the QR regime.  

Beyond this point, the Moon's eccentricity rapidly decreases due to lunar tides, and its semimajor axis accordingly re-starts its expansion. The evection resonance is now  interior to the Moon, and its position continues to move inward as Earth's spin decreases. Therefore the relative positions of the Moon and the resonance  diverge with time and evection no longer affects the lunar orbit. Further evolution occurs at a nearly constant Earth-Moon AM via standard tidal evolution. In the simulation shown in Fig \ref{fig:EvolutionFastTides}, the final AM is close to current AM of the Earth-Moon system (grey area in Fig. \ref{fig:EvolutionFastTides}-d). As described in section \ref{sec:ConstTidalParameters}, the final AM depends on when escape from QR occurs, which depends on tidal parameters.

Whenever escape from proper evection resonance places the trajectory in the outer circulating region (i.e., to the high-$e$ side of evection resonance
), a QR-regime follows. This occurs in $100\%$ of cases in which $A\gtrsim$ few hundred, because for these cases the resonance is in the shallow resonant regime (Fig. \ref{fig:LevelCurves}a-b) at the time of escape and only the outer circulating region exists. For lower $A$, the deep resonance regime (Fig. \ref{fig:LevelCurves}c-d) applies at the time of escape, and it is possible for the Moon to also escape from evection proper to the low eccentricity side of the librating region (i.e., into the central circulating region on the phase diagram; see Fig. S4 for an example of this evolution). In this case, the lunar orbit is exterior to the position of evection, $a^{\prime}_{\rm res}$, upon escape, and no QR regime occurs, leaving the Earth-Moon AM at a very high value. Two cases out of ten simulation performed for the combination of $A$ and $t_T$ in Fig. \ref{fig:EvolutionFastTides} and Fig. S4, but with different initial resonance angles, $\varphi(0)$, experienced this non-QR behaviour. The probability of exiting the resonance above/below the stable eccentricity varies according to the tidal parameters, as these affect the shape and relative area of the non-QR circulating region (Fig. \ref{fig:LevelCurves}c-d). In general, as the relative tidal strength $A$ decreases, the maximum critical eccentricity before orbit contraction increases (see next section), and escape may increasingly occur into the non-QR region (i.e., the central circulating region). However we find that even for $A\sim$unity, escape usually occurs into the QR region.

\subsection{Dependence on tidal parameters}\label{sec:ConstTidalParameters}

Evection resonance and the subsequent QR can remove adequate AM to be consistent with the current Earth-Moon (grey area in Fig. \ref{fig:EvolutionFastTides}). In order to estimate the likelihood of such potentially successful cases (final AM $\sim L_{\rm EM}$), we performed simulations across a large range of tidal parameters, $5\leq A\leq 10^4$ and $20\leq Q_{\rm eff}/{k_{2\oplus}}\leq10^5$. 

Initial capture into evection resonance requires that the time for the upward movement of the stationary point, $e_s/|\dot{e_s}|$, is longer than the period of libration around that point ($P_{\rm {lib}}=\pi/4\Omega_\odot t_T \sqrt{5\alpha e^2_*}$, \citealp{ward2020analytical}). Therefore, there is a minimum eccentricity for capture $e_{\rm crit}\propto\sqrt{1/t_T}$. For large $A$ values, the eccentricity derivative during the tidal evolution towards evection resonance is negative  ($|\dot{e}_\oplus|<|\dot{e}_m|$). Eventually, for strong enough lunar tides the eccentricity at the evection resonance position is smaller than $e_{\rm crit}$, preventing capture into evection resonance (e.g., for $Q_{\rm eff}/k_{2\oplus}=42$ capture requires $A\lesssim40$; see Fig. S5-a). Furthermore, for fast terrestrial tidal rates (small values of $t_T$, or equivalently small $Q_{\rm eff}/k_{2\oplus}$) the minimum required eccentricity, $e_{\rm crit}$, increases, precluding capture into evection (e.g., for $A=50$ capture occurs when $Q_{\rm eff}/k_{2\oplus}\gtrsim210$; see Fig. S5-b). Overall, for large values of $A$ and fast terrestrial tidal rates, the Moon exits proper evection before the resonance is fully developed, and Earth-Moon AM remains high.

\begin{figure} 

     \centering
     \includegraphics[width=0.8\linewidth]{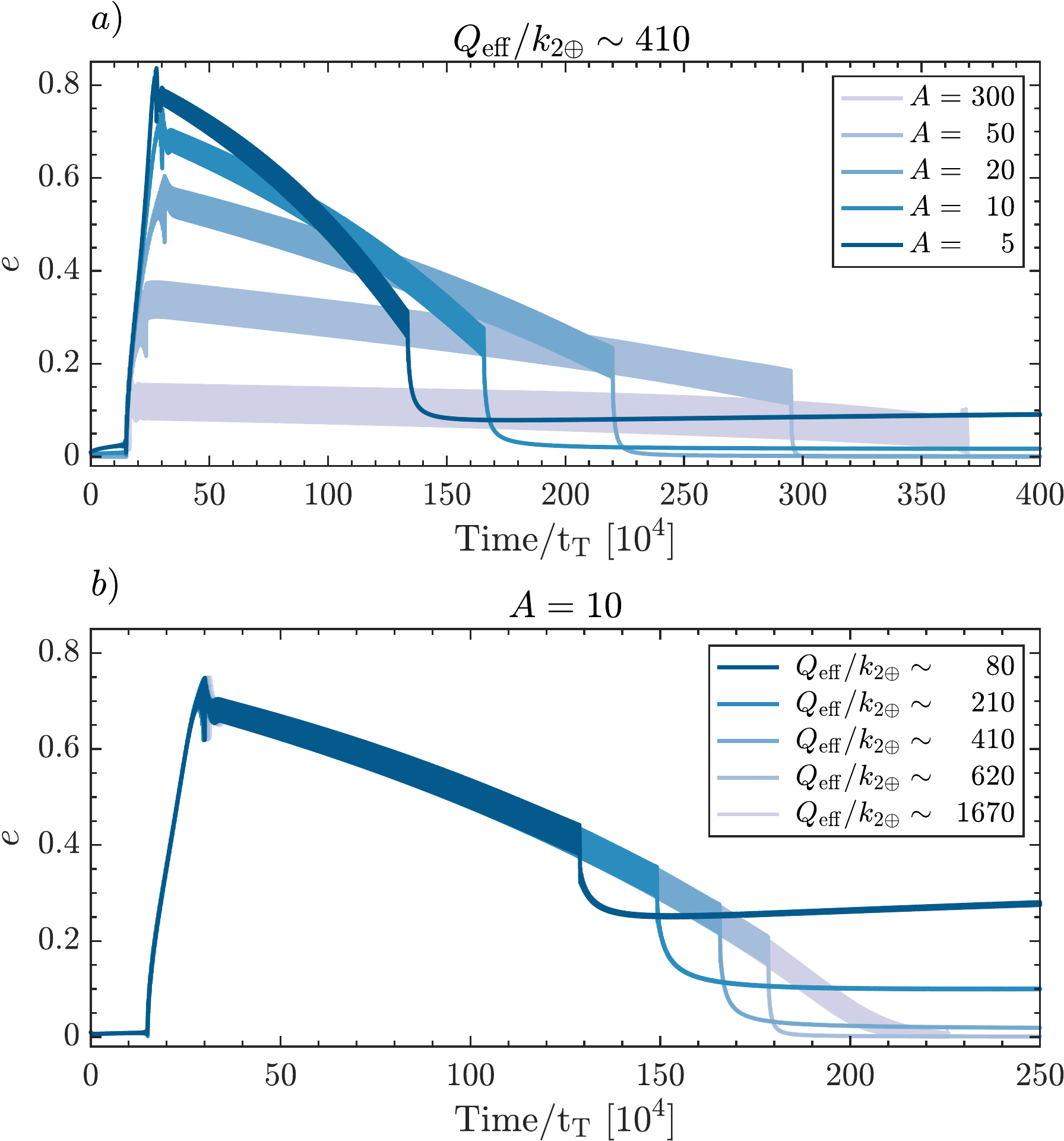}

     \caption{\justifying Eccentricity evolution with different a) relative tidal strength values, $A$ (assuming terrestrial tidal parameters $Q_{\rm eff}/k_{2\oplus}\sim410$); b) terrestrial tidal parameters, $Q_{\rm eff}/k_{2\oplus}$ (assuming $A=10$). The time is normalized by the tidal timescale rate, $t_T$ (see text) and assumes a non-synchronous lunar rotation. For larger values of $A$ with fixed $Q_{\rm eff}/k_{2\oplus}$, the maximum eccentricity excited during evection resonance decreases.  When QR escape occurs, the evolution of the Moon's eccentricity shifts from an oscillating behavior whose value is controlled by evection, to a standard evolution controlled by tides. For larger values of $Q_{\rm eff}/k_{2\oplus}$ with fixed $A$, the exit from the QR regime occurs later and at smaller eccentricities. With $A=10$ and $Q_{\rm eff}/k_{2\oplus}\gtrsim1670$, the co-synchronous state in which the lunar month equals the Earth's day is achieved.}
      \label{fig:DiffA}
 \end{figure}

For cases that result in initial capture into evection resonance, Fig. \ref{fig:DiffA} shows the evolution of the eccentricity given different relative strength values, $A$, and different absolute tidal rates (represented by $Q_{\rm eff}/k_{2\oplus}$). For cases with $A<4$, the eccentricity excitation is large, and the Moon's perigee becomes smaller than a high-$e$ tidal  disruption boundary, $q^{\prime}<2R_\oplus$ \citep{sridhar1992tidal}, therefore we considered $A\geq5$. For increasing tidal dissipation inside the Moon (higher $A$), the maximum eccentricity reached is smaller (Fig. \ref{fig:DiffA}-a), because stronger lunar tides overcome Earth tides at a smaller eccentricity.


The angular momentum at the end of proper evection is $L^{\prime}_{\rm esc}\approx  a^{\prime7/4}_{\rm esc}(1-e_{\rm esc}^2)/\Lambda+\gamma \sqrt{a^{\prime}_{\rm esc}(1-e_{\rm esc}^2)}$, where we assumed that the lunar spin contribution is small and that during resonance $\dot{\varphi}\approx0$, hence $s^\prime \approx a^{\prime7/4}(1-e^2)/\Lambda$. For lower $A$ values, $a^{\prime}_{\rm esc}(1-e_{\rm esc}^2)$ decreases (see Fig. S6), so that $L^{\prime}_{\rm esc}$ is smaller, therefore more AM is removed during evection proper. However for low $A$ and high-$e$ evolutions, the assumption here that the lunar time delay is constant with time is not a good approximation. \cite{tian2017coupled} showed that tidal heating in the Moon during such high-$e$ evolutions would lead to melting and increased lunar dissipation, ultimately causing early exit from evection (we return to this point in Section 4).

For $A\gtrsim400$, the overall excitation of eccentricity and semimajor axis during evection resonance is minimal, and there are no real solutions for eqn. (\ref{eq:esx}), hence while the stable stationary eccentricity, $e_s$, is small, the saddle stationary point is $e_{sx}=0$ (this regime is referred to as the shallow resonance; Fig. \ref{fig:LevelCurves}a-b). In this case, the central circulating region of clockwise motion (Fig. \ref{fig:LevelCurves}c-d) does not exist, and escape from proper evection always occurs on the QR side. 

To explain the differences in the timing of QR escape (Fig. \ref{fig:DiffA}b), we combine the circulation timescale (eqn. \ref{eq:libP}) and the crossing timescale (eqn. \ref{eq:Tcross}), and define the minimum distance from the separatrix needed to maintain the system in QR:
\begin{linenomath*}
 \begin{equation}
     e-e_{sx}=\frac{\pi|\dot{e}_\oplus+\dot{e}_m|}{2t_T\Omega_{\odot}\sqrt{5\alpha}e_*}
     \label{eq:e-esx}
 \end{equation}
\end{linenomath*}
Hence, with slower terrestrial tidal rates (larger $t_T$ or $Q_{\rm eff}/k_{2\oplus}$) the eccentricity more closely approaches the separatrix before exiting the QR regime, resulting in more AM removal. For slow enough terrestrial tidal rates, QR drives the system all the way to the dual synchronous state (light blue curve in Fig. \ref{fig:DiffA}-b; see also Fig. S7), assuming that the tidal parameters $\Delta t$ and $A$ remain unchanged during the evolution. 

Moreover, because the eccentricity excitation is shallower as $A$ increases (Fig. \ref{fig:DiffA}-a), the tidal derivative magnitude, $|\dot{e}_\oplus+\dot{e}_m|$, is lower as well. Therefore, with larger $A$ values, the exit from QR regime occurs at a lower eccentricity as well, although the dependence is less substantial.

\begin{figure}
    \centering
     \includegraphics[width=0.8\linewidth]{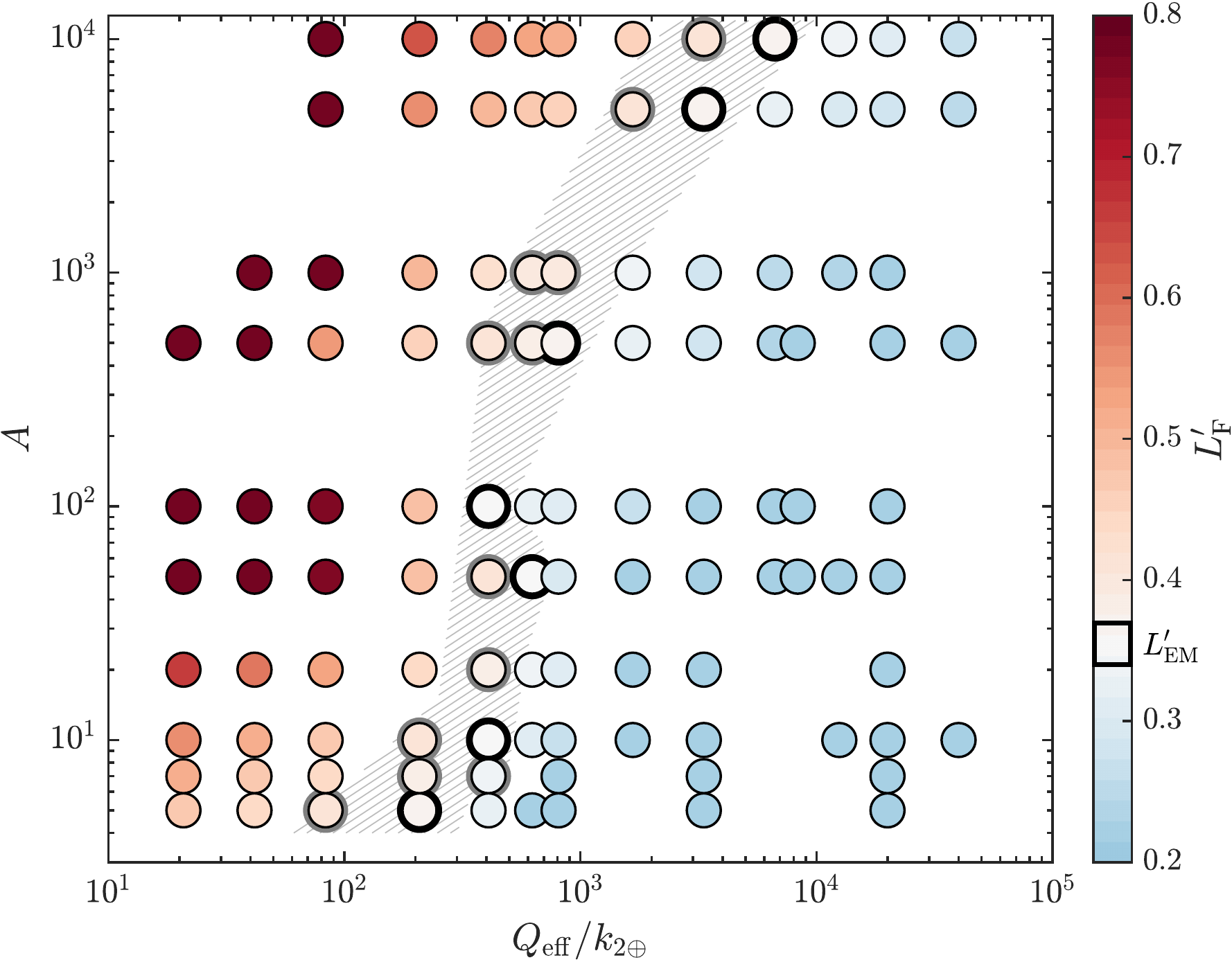}
    \caption{\justifying The final AM as a function of the initial terrestrial tidal parameters, $Q_{\rm eff}/k_{2\oplus}$, and the relative tidal strength factor, $A$, for the cases in which the Moon's rotation evolves solely due to tides (i.e., non-synchronous lunar spin for eccentric orbits) and the tidal parameters are assumed to be constant with time. The initial AM for these simulations is $L^\prime\sim0.77$ ($2.2L_{\rm EM}$), and the co-synchronous state has $L^\prime\sim0.23$ ($\sim0.6 L_{\rm EM}$), which is the minimum final AM. For each combination of $A$ and $Q_{\rm eff}/k_{2\oplus}$, ten initial $\varphi(0)$ values were simulated. The dispersion of the final AM in each combination is mostly dependent on whether the system exits the evection proper regime on the QR side or not. In the figure we show the minimum final AM in each combination, obtained in the simulations that exited on the QR side. We define the range of successful simulations with $0.99-1.07\, L_{\rm EM}$ (black circles), and in addition we highlight cases whose final AM lie outside the successful range but within the range $0.97 - 1.21 L_{\rm EM}$ (grey circles). The former accounts for AM alterations due to late accretion \citep{bottke2010stochastic} and solar tides \citep{canup2004simulations}. The later range was arbitrarily chosen to reflect approximately 3 times more AM alteration than expected from these subsequent processes.}
     \label{fig:AVsQColor}
 \end{figure}

In general, the two tidal parameters,  $A$ and $Q_{\rm eff}/k_{2\oplus}$, have opposing effects on the final AM. As $A$ increases for fixed $Q_{\rm eff}/k_{2\oplus}$, the final AM increases (less AM is removed), because the eccentricity excitation is lower.  However, as $Q_{\rm eff}/k_{2\oplus}$ increases for fixed $A$, the final AM decreases (more AM is removed) because the system stays in QR longer and exits QR at a lower eccentricity. Hence, parameters that could produce a final AM of $\sim1\,L_{\rm EM}$ are correlated and follow the narrow inclined shaded band in Fig. \ref{fig:AVsQColor}.

Although the $A$ and $Q_{\rm eff}/k_{2\oplus}$ values appropriate during the Moon's encounter with evection are uncertain, a compelling argument has been made by \cite{zahnle2015tethered} that both would have been large. The Moon-forming impact would have likely heated Earth's mantle to an initially molten, fluid-like state \citep{Nakajima2015286}, implying $Q\sim10^3$ to $10^5$, based on analogy to gas giant planets (e.g., \citealt{lainey2009strong,lainey2012strong} infer $Q/k_2\approx9\times10^4$ for Jupiter and $\approx 4 \times 10^3$ for Saturn, respectively). As Earth's mantle later began to solidify, its viscosity would have increased by orders-of-magnitude, causing greatly increased tidal dissipation and low $Q$ values of order unity. However, there is a limit to how soon the Earth's mantle could have transitioned to this low-$Q$ state, due to the volatile-rich atmosphere that would have persisted for a few Myr and controlled the rate of planet cooling. Consider a mantle that has started to solidify. If tidal heating exceeds what can be accommodated via atmospheric cooling, mantle temperatures would increase and the mantle crystal fraction would decrease, causing the mantle viscosity to decrease and with it tidal dissipation/heating.  Conversely, if tidal dissipation in the mantle generated less heat than accommodated via radiative cooling, additional mantle freezing would occur, causing the mantle viscosity and tidal dissipation/heating to increase. \cite{zahnle2015tethered} argue that this feedback would cause Earth's initial tidal dissipation rate to be regulated by the modest cooling rate of its early dense atmosphere, yielding large initial terrestrial $Q$ values that are similar to those inferred for giant planets when the Moon encountered evection.

Such slow orbital expansion rates would promote initial capture into evection resonance, and imply that the Moon reaches the evection resonance point, $a^{\prime}_{\rm res}$, in  $\sim$ few $\times 10^5$ to few $ \times 10^6\ \rm{yr}$. Compared to this timescale, the lunar crustal formation time is fast ($\sim10^3$ yr for solidification of $80\%$ of the magma ocean and lid formation; \citealp{Elkins2011}). In a partially solid early Moon, tidal dissipation is expected to be relatively high (i.e., large $\Delta t_m$), comparable to or higher than in the current Moon. Therefore the relative tidal strength, $A$, during evection would be high as well. Assuming lunar dissipation similar to the current value, \cite{zahnle2015tethered} estimated  $A_Q\sim10^{2}-10^{4}$ when the Moon encounters evection (note that their results are described by the parameter $A_{Q}$ which is consistent with the constant-$Q$ model, and is a factor of $n/(s-n)\sim \mathcal{O}(10^{-1})$ smaller than the $A$ used here from eqn. \ref{eq:A}). The likely presence of a lunar magma ocean when evection resonance occurs would increase tidal dissipation in the early Moon compared to that in the current Moon, hence increasing $A$ even further. 

Assuming that the terrestrial and lunar tidal parameters are constant throughout the evolution (i.e., that $\Delta t$ and $A$ are constant; see section \ref{sec:VariableTau} for cases in which these tidal parameters vary with time as Earth cools), the co-synchronous state is the most common final state across the range  $10^3<Q_{\rm eff}/k_{2\oplus}<10^5$ and $10^2<A<10^{4}$ (blue markers in Fig. \ref{fig:AVsQColor}). A small part of this range, e.g., for $A\sim10^4$ and $Q_{\rm eff}/k_{2\oplus}\sim4\cdot10^3$, is consistent with the final AM of $1\,L_{\rm EM}$ (markers highlighted by the black circles in Fig. \ref{fig:AVsQColor}). 

\subsection{Synchronous lunar spin}
The above simulations assume that the lunar spin evolves solely due lunar tidal torques (eqn. \ref{eq:sm}). These produce an approximately synchronous rotation for nearly circular orbits, but yield $s_m>n$ for eccentric orbits.
However, a tri-axial lunar shape (frozen in during lunar cooling; e.g., \citealp{garrick2006evidence}) can maintain a synchronous spin even when the orbit is substantially eccentric if the figure torque is larger than the tidal torque \citep{goldreich1966spin}. In this case, $s^{\prime}_m=n^{\prime}$ (where $n^{\prime}\equiv a^{\prime-3/2}$ is the normalized lunar mean motion), and the Moon's spin is altered by both lunar tides and the permanent figure torque.  We note that although the lunar lag angle when the Moon is in synchronous rotation is zero ($\delta_m=(s_m-n)\Delta t_m=0$), dissipation of energy still occurs with Mignard tides for nonzero eccentricity, due to variations in the magnitude of the tide from periapse to apopase \citep{burns1986evolution}. Thus, even with $s_m = n$, there are changes in the Moon's orbit due to lunar tides if its orbit is non-circular (see also Text S1).

For a Moon with principal moments of inertia $C_m\ge B_m\ge A_m$, the orbit-averaged value of the permanent figure torque (\textit{pf}) is (e.g., \citealp{goldreich1966spin}):
\begin{linenomath*}
\begin{equation}
    \langle T_{pf}\rangle = -\frac{3}{2}\Omega_{\oplus}^2a^{\prime-3/2}(B_m-A_m)\mathcal{H}(e)\sin{2\psi_0}=(\Omega_\oplus/t_T) C_m\dot{s}^{\prime}_{m,pf}
    \label{eq:T_pf}
\end{equation}
\end{linenomath*}
where $\mathcal{H}(e)=1-5e^2/2+13e^4/16$, is the so-called Hansen polynomial, $\psi_0$ is the angle between the long axis of the Moon and the Earth-Moon line at perigee, and $\dot{s}^{\prime}_{m,pf}$ is the normalized resulting change in the lunar spin rate. This torque leads to additional contributions to $\dot{a}^{\prime}$ and $\dot{e}$, which we parameterize as \citep{ward2020analytical}:

\begin{linenomath*}
\begin{equation}
     \frac{\dot{a}^{\prime}_{pf}}{a^{\prime}}=-f_{pf}\left[\frac{\dot{a}^{\prime}_m}{a^{\prime}}-\frac{2e\dot{e}_m}{1-e^2} \right]
          \label{eq:PF_additions_a}
 \end{equation}
 \begin{equation}
     \dot{e}_{pf}=\frac{g_{pf}}{2e}\left[\frac{\dot{a}^{\prime}_m}{a^{\prime}}-\frac{2e\dot{e}_m}{1-e^2} \right]
     \label{eq:PF_additions_e}
 \end{equation}
\end{linenomath*}
where $f_{pf}=\sqrt{1-e^2}$ and $g_{pf}=(1-e^2)(1-\sqrt{1-e^2})$; to second order in $e$, the eqn. \ref{eq:PF_additions_a} and \ref{eq:PF_additions_e} expressions are comparable to those used in prior studies (see Text S1).

Synchronous rotation can be maintained if $\dot{s}^{\prime}_{m,pf}$ approximately balances that due to lunar tides, $\dot{s}^{\prime}_m$, with the angle $\psi_0$ adopting the needed value. However $|\sin{2\psi_0}|$ has a maximum value of unity, so that there is a minimum value of $(B_m-A_m)/C_m$ needed for $\dot{s}^{\prime}_{m,pf}\approx|\dot{s}^{\prime}_m|$
\begin{linenomath*}
\begin{equation}
\begin{aligned}
    \frac{B_m-A_m}{C_m}&>\left|\frac{\gamma}{3\kappa}
    \left(\frac{a^{\prime7/2}}{\Omega_\oplus t_T}\right)
    \frac{\sqrt{1-e^2}}{\mathcal{H}(e)}
    \left(\frac{\dot{a}^{\prime}_m}{a^{\prime}}-\frac{2e\dot{e}}{1-e^2}\right)\right| \\
   &>\left|4\times10^{-4}\left( \frac{k_m\Delta t_m}{\rm{4\ min}}\right)\left( \frac{7}{a^{\prime}}\right)^{9/2}\frac{\sqrt{1-e^2}}{\mathcal{H}(e)}\left[f_1-f_2-\frac{e^2}{1-e^2}(g_1-g_2)\right]\right|
\end{aligned}
\end{equation}
\end{linenomath*}
where $k_m\Delta t_m\approx4\ \rm{min}$ for the current Moon \citep{williams2015tides}. If this condition is not met, non-synchronous motion will result. For a low-eccentricity orbit, a value comparable to that for the current Moon's shape, $(B_m-A_m)/C_m=2.28\times10^{-4}$ \citep{yoder1995astrometric}, could maintain synchronous lock at the time of evection encounter. However, as $e$ increases, a progressively larger value is needed; e.g., for $e=0.5$, $(B_m-A_m)/C_m>10^{-2}$ is required, which may be difficult to maintain. Thus, a synchronous rotation model would seem most aptly applied to high $A$ cases (in which the lunar eccentricity remains modest), while non-synchronous rotation may be more likely for low $A$ cases in which $e$ achieves large values.

\begin{figure}
     \centering
     \includegraphics[width=0.8\linewidth]{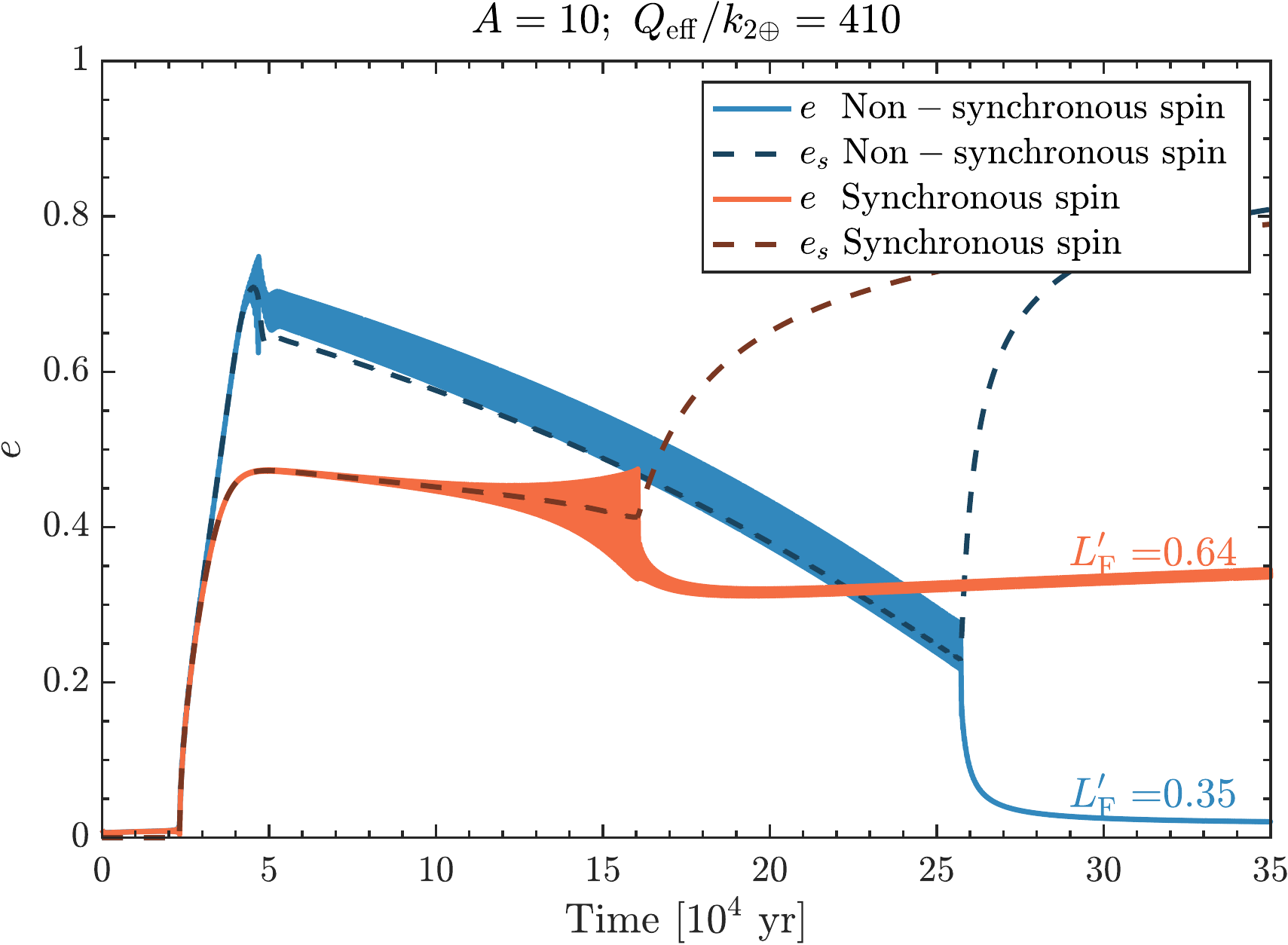}
     \caption{\justifying Evolution of the eccentricity (solid lines) and stable stationary eccentricity (dashed lines) for non-synchronous (blue) and synchronous lunar rotation cases (orange). The latter includes the effect of permanent figure torques that are assumed to maintain the synchronous rotation (see text for details). The synchronous case experiences a smaller eccentricity excitation, and its libration amplitude growth while in resonance is more gradual, allowing the Moon to remain in evection proper for longer (until $t\sim16\times10^4$ yr) compared to the non-synchronous case (that exits proper resonance at $t\sim4.5\times10^4$ yr). For all simulated cases with $A\leq100$, the exit from evection proper for the synchronous case occurs below the stable eccentricity as is the case here, so that there is no QR regime. However, as discussed in the text it is not clear that synchronous rotation could be maintained for these low-$A$ cases that lead to substantial lunar eccentricities. The final normalized AM is listed above each case.}
     \label{fig:PermFigVsNot}
 \end{figure}

Fig. \ref{fig:PermFigVsNot} shows the eccentricity evolution for a synchronous case (orange line) compared to a non-synchronous case (blue line). The maximum eccentricity before the orbital contraction is lower in the synchronous case, and the libration amplitude growth is more gradual, therefore the system remains in evection proper longer compared to the non-synchronous case (as predicted by \citealp{ward2020analytical}). 

\begin{figure}
     \centering
     \includegraphics[width=0.8\linewidth]{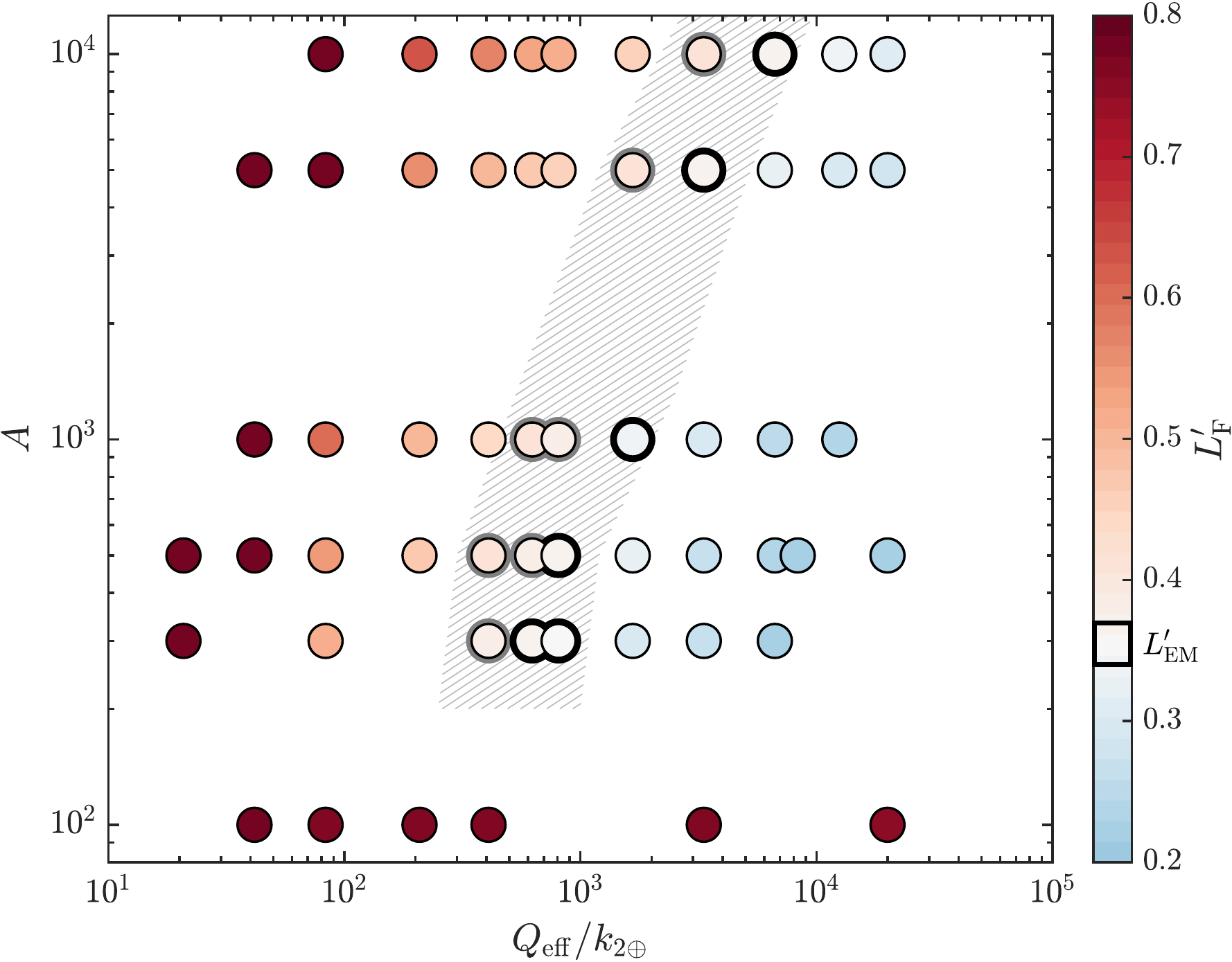}
    \caption{\justifying Similar caption to Fig. \ref{fig:AVsQColor} for cases that include the permanent figure torque (synchronous lunar spin). For tested simulations with $A\leq100$, the Moon did not enter the QR regime, hence the AM remains high.}
     \label{fig:AVsQColorPF}
 \end{figure}

For the cases we have tested with synchronous rotation and $A\le 100$, exit from evection proper always occurs towards the central non-resonant region (or below the stable eccentricity; dashed line in Fig. \ref{fig:PermFigVsNot}), and no quasi-resonant evolution occurs. The permanent figure torque causes an enhanced damping of the Moon's eccentricity compared to that due to lunar tides alone (i.e., eqn. \ref{eq:PF_additions_e} is always negative), with the magnitude of the $\dot{e}_{pf}$ term depending on both $A$ and the $f$ and $g$ functions, which in turn depend on $e$.  Because the eccentricity as the Moon exists evection resonance increases as $A$ decreases, the $\dot{e}_{pf}$ term actually becomes stronger as $A$ decreases.  For low enough $A$, when resonant control of $e$ ends, the $\dot{e}_{pf}$ term is strong enough to pull the trajectory directly downward on the level curve diagram into the central region, precluding establishment of the QR for synchronous rotation cases with $A\leq100$ (with little change in the AM; dark red markers in Fig. \ref{fig:AVsQColorPF} each representing 10 simulations with different initial $\varphi$). However, per above, i) it is not clear that low-$A$/high-$e$ cases would be consistent with synchronous rotation and ii) low-$A$/high-$e$ cases would likely rapidly transition to high-$A$, due to the effects of tidal heating within the Moon for high $e$ \citep{tian2017coupled}.   
  
For cases with $A\gtrsim300$, the eccentricity excitation is small and the region below the separatrix does not exist ($e_{sx}=0$). Hence, similar to the non-synchronous cases, exit from evection proper always occurs towards the QR region. For low eccentricities associated with high-$A$ cases, the additional permanent figure terms are small ($f_{pf}\approx 1$ and $g_{pf}\approx0$), therefore, the  timing of QR escape (governed by eqn. \ref{eq:e-esx}) and the final Earth-Moon AM (Fig. \ref{fig:AVsQColorPF}) is similar to the non-synchronous case.

\subsection{Effects of planetary cooling: $Q(t)$}\label{sec:VariableTau}

The simulations above consider constant tidal parameters ($\Delta t$ and $A$) throughout the evolution. Earth's early opaque atmosphere would have kept its mantle molten and its $Q$ very large for $2$ to $5$ Myr after a high-AM giant impact \citep{zahnle2015tethered}. After this time the mantle viscosity increases dramatically over the next few Myrs as the mantle begins to freeze, which causes $Q$ to plummet to values $\sim$ unity \citep{zahnle2015tethered}. The time needed for the Earth's mantle to begin to solidify is typically somewhat shorter than the total time required for  passage through evection proper and QR (e.g., for $Q_{\rm eff}/k_{2\oplus}\sim10^4$ and $A=5000$, QR escape occurs at $\sim20\ \rm{Myr}$). Therefore, while large $Q$ likely applied when the Moon was captured into evection resonance, during the QR regime it may have decreased dramatically.  

To estimate the effect of planetary cooling, we performed additional simulations with a varying terrestrial time lag, $\Delta t$. We assume that Earth's tidal dissipation factor, $Q$, evolves  according to \cite{zahnle2015tethered}, Fig. 8, which depicts Earth's thermal evolution after a high-AM impact. We consider either a peak value of $(Q/k_{2\oplus})\approx 7 \times 10^3$ (solid curves in \citealp{zahnle2015tethered}, their Fig. 8) or $(Q/k_{2\oplus})\approx 7 \times 10^4$ (dashed curves in  \citealp{zahnle2015tethered}, their Fig. 8).

For each time step, $i$, we calculate Earth's time lag according to:
\begin{linenomath*}
\begin{equation}
    \Delta t_i=\frac{1}{2\Omega_{\oplus} (s^\prime_{i-1}-n^\prime_{i-1})Q_{{\rm Zahnle},i}}
    \label{eq:DeltaTVar}
\end{equation}
\end{linenomath*}
where $s^{\prime}$ and $n^{\prime}$ are sampled from previous time step, $i-1$, and $Q_{{\rm Zahnle},i}$ evolves with time per the \citeauthor{zahnle2015tethered} model.
The relative strength of lunar tides compared with Earth tides is calculated using:
\begin{linenomath*}
\begin{equation}
    A_i=A_{\rm today} \left(\frac{\Delta t_{\rm today}}{\Delta t_i}\right) \left(\frac{k_{2,\oplus}}{k_{2, \rm{fluid}}}\right) \left(\frac{\Delta t_m}{\Delta t_{m,\rm{today}}}\right)
    \label{eq:AVar}
\end{equation}
\end{linenomath*}

where $\Delta t_{\rm today}$ [$\Delta t_{m,\rm today}$] is Earth's [lunar] current time lag, we set $k_{2,\rm{fluid}} = 1.5$ as in \cite{zahnle2015tethered}, and \change{$A_{\rm today}\sim16$}{$A_{\rm today}\sim13$} is the current ratio (assuming values of $k_{2m}=0.024$, $\Delta t_{m,\rm {today}}= 165\ \rm {min}$, \citealt{williams2015tides}; \change{$k_{2\oplus}=0.25$}{$k_{2\oplus}=0.30$} and $\Delta t_{\rm today}=10\ \rm{min}$, \citealt{murray1999solar}). For a large terrestrial $Q$, the Moon evolves outward slowly and has time to partially solidify by the time evection resonance is encountered, leading to strong dissipation in the Moon comparable to or greater than in the Moon today \citep{zahnle2015tethered}. During initial evection resonance capture, $A$ will be high, therefore the eccentricity excitation is low, and minimal heating inside the Moon due to evection is expected  \citep{tian2017coupled}. Accordingly, we assume that the lunar lag time is constant throughout the evolution. The current lunar time delay is due to monthly dissipation as the Moon is synchronously rotating, but the Moon's rotation state and time delay during evection resonance and QR are uncertain. To account for the uncertainty on the appropriate lunar time lag during the evolution (e.g., the last $20\%$ of the magma ocean solidification requires $10\ \rm{Myrs}$ due to formation of anorthositic crust; \citealp{Elkins2011}), we consider different values for $f_m\equiv \Delta t_m/\Delta t_{m,\rm{today}}$. Values of $f_m>1$ imply more tidal dissipation in the early Moon than in the Moon today, representing possible enhanced dissipation due to an early lunar magma ocean. 

\begin{figure}
     \centering
     \vspace{-2cm}
     \includegraphics[width=0.85\linewidth]{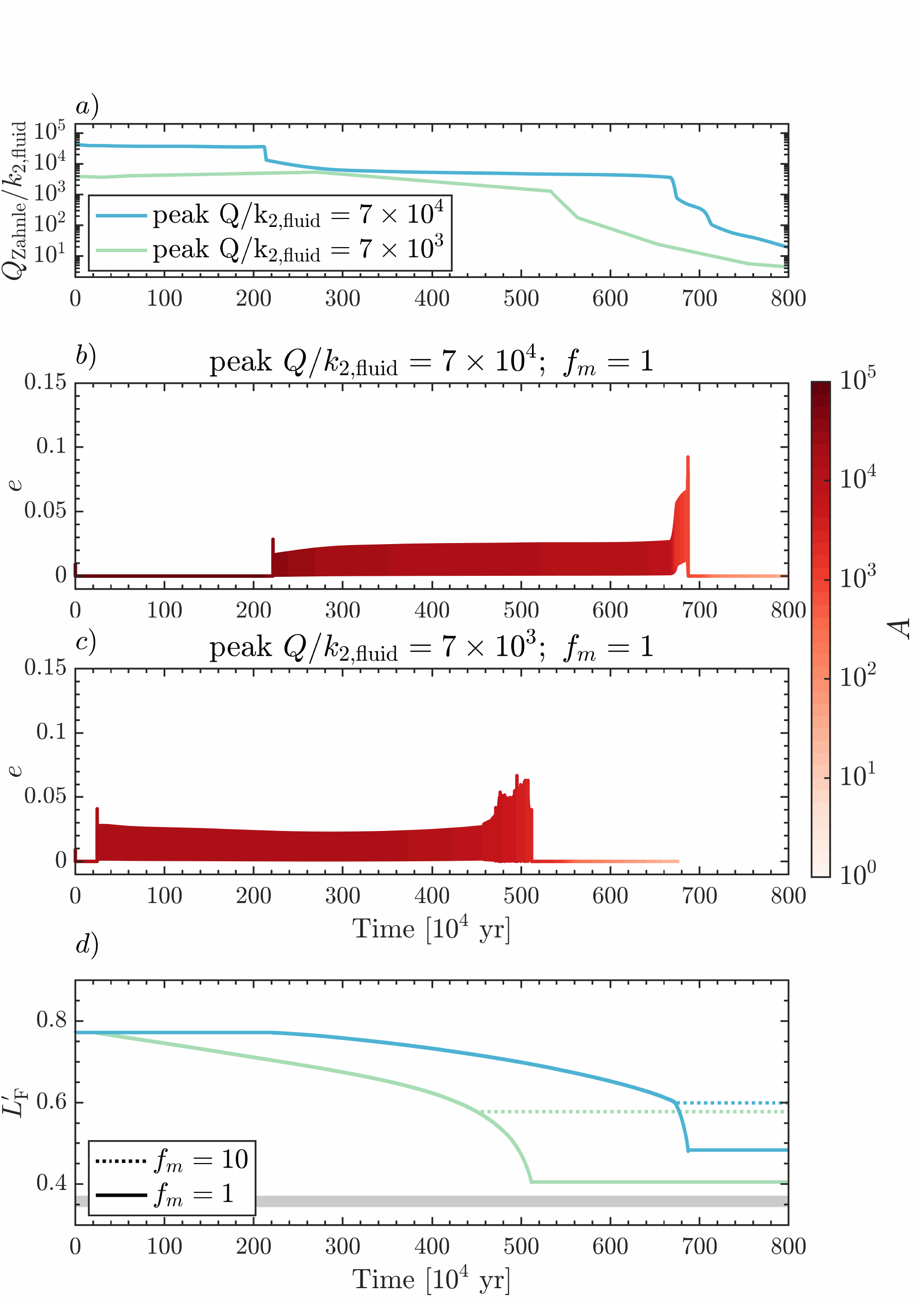}
     \thisfloatpagestyle{empty}
     \caption{\justifying Evolution through evection resonance and the subsequent QR-regime with time dependent tidal dissipation in the Earth and a non-synchronous lunar rotation. a) Time-dependent terrestrial tidal dissipation factor model by \cite{zahnle2015tethered} after a high-AM Moon forming impact (the low and middle curve in their Fig. 8). The higher (blue) and lower (green) peak terrestrial $(Q/k_{2,\rm fluid})$ values are comparable to values inferred from astrometry for Jupiter and Saturn \citep{lainey2009strong,lainey2012strong}. Lunar eccentricity during passage through evection assuming a) a peak $Q/k_{2,\rm fluid} = 7 \times 10^4$ and $f_m=1$; b)  a peak $Q/k_{2,\rm fluid} = 7 \times 10^3$ and $f_m=1$.  The evolution of the relative tidal strength, $A$, is indicated by the red colors. d) Normalized AM evolution of the Earth-Moon system for $f_m=1$ (i.e., with lunar tidal parameters comparable to those in the current Moon; solid line) and $f_m=10$ (corresponding to enhanced dissipation in the early Moon due to the presence of a magma ocean; dashed line), for the higher (blue) and lower (green) peak terrestrial $Q/k_{2,\rm fluid}$ cases. The horizontal grey area  represents values consistent with the current Earth-Moon, accounting for later AM change due to solar tides and late accretion impacts \citep{canup2004simulations,bottke2010stochastic}.}
    \label{fig:VariableTauLow}
 \end{figure}

Fig. \ref{fig:VariableTauLow}  shows the lunar eccentricity evolution with peak $(Q_{\rm Zahnle}/k_{2,\rm{fluid}})$ values comparable to those inferred for Jupiter and Saturn from astrometry \citep{lainey2009strong,lainey2012strong}, and assuming non-synchronous lunar rotation. The Moon is captured into evection proper at $2.1\ \rm{Myr}$ [$0.2\ \rm{Myr}$] for the higher [lower] initial dissipation factor, followed by escape and a QR regime. As the tidal parameter, $A$, begins to rapidly decrease due to Earth's mantle cooling (lighter colors in Fig. \ref{fig:VariableTauLow} - b, c), the eccentricity increases during the QR regime as lunar tides become relatively less efficient compared to the strengthening Earth tides. For the slower tidal evolution (higher initial dissipation factor; blue curve in Fig. \ref{fig:VariableTauLow}-d), most of the AM removal occurs after the mantle begins to solidify and $(Q_{\rm Zahnle}/{k_{2,\rm fluid}})$ drops ($\sim 6.5\ \rm{Myr}$). QR escape occurs at $\sim 6.9\ \rm{Myr}$ when the timescale of circulation is comparable to the timescale of tidal changes (Fig. S8). In contrast to the constant time lag case (section \ref{sec:ConstTidalParameters}), during Earth's cooling, the stable eccentricity gradually grows (level curves evolve outward  from shallow resonance, Fig. \ref{fig:LevelCurves}-a, to deep resonance, Fig. \ref{fig:LevelCurves}-d). Eventually Earth's tidal strength increases (due to the sharp decrease in Earth's tidal dissipation factor; Fig. \ref{fig:VariableTauLow}-a) at a rate that is comparable to the period of lunar circulation in QR ($t_{\rm cross}\sim P_{\rm circ}$; see Fig. S8), and exit from QR occurs. In both $(Q_{\rm Zahnle}/k_{2,\rm{fluid}})$ evolutions the final system is left with too much AM for consistency with the Earth-Moon (Fig. \ref{fig:VariableTauLow}-d). The mismatch worsens if, as is expected, early lunar dissipation was higher than in the current Moon ($f_m>1$; dashed lines in Fig. \ref{fig:VariableTauLow}-d).

\section{Conclusions and discussion}
In this work we modeled the lunar tidal \remove{evolution }evolution during its encounter with the evection resonance. The Moon tidally recedes away from its accretion location near the Roche limit, while its perigee precession period increases. The Moon encounters evection resonance when the precession period equals one year. During evection proper, the eccentricity increases and the Moon librates around the stable stationary point, $\varphi\sim\pm\pi/2$ (representing the angle between the Moon's perigee and the Sun). As the eccentricity increases, lunar tides (governed by the tidal parameter $A$) become stronger, and eventually cause the lunar orbit to contract. During resonance, the solar net torque controls the Moon's eccentricity, while tides still control the evolution of the Moon's semimajor axis, and the balance between the lunar orbital angular momentum (AM) and the spins of the Earth and Moon (eqn. \ref{eq:L}) can no longer be maintained. As such, AM is removed from the Earth-Moon system and transferred to the Earth's heliocentric orbit. Near the turn-around point, when the Moon's orbit begins to contract, the libration amplitude increasess and leads to escape from proper resonance. This is consistent with escape timing seen in \cite{touma1998resonances} with the Mignard model for initial AM values of $\sim L_{\rm EM}$, and with analytic predictions of libration growth in \cite{ward2020analytical}.

If the lunar spin evolves only due to tides, once the orbit contracts and exits evection proper, it typically enters a quasi-resonance (QR) regime. Although the Moon is not librating around a stationary point and thus is not in formal resonance, evection still regulates the eccentricity evolution during QR and it is during the QR-regime that substantial AM can be removed from the Earth-Moon pair. The QR regime is reminiscent of the limit cycle found by \cite{Wisdom2015138}, and thus is probably not an artifact of the chosen tidal model (Mignard vs. constant-$Q$ tides). We find that for Mignard tides, the final AM depends on the timing of QR escape, and prolonged occupation of the QR state may result in a co-synchronous state, inconsistent with the Moon's long-term survival. 

If the Moon had a permanent frozen figure that maintains synchronous rotation even when its orbit is eccentric, additional terms to the eccentricity and semimajor axis evolution apply. In this case, for relatively weak lunar tides ($A<300$), the Moon always exits the proper evection regime on the non-QR side of the resonance, hence overall only minimal  AM is removed. In addition, for very low $A$, tidal heating in the Moon would likely lead to rapid escape from resonance (\citealp{tian2017coupled}, see below). For stronger lunar tides ($A>300$), the eccentricity excitation is small and the additional permanent figure terms are negligible, yielding similar results as the non-synchronous case.

For constant tidal parameters, our results imply that only for a narrow range in both $A$ and $Q_{\rm eff}/k_{2\oplus}$ would a post-giant impact system with about twice the AM of the Earth-Moon system have its AM appropriately reduced to $\sim L_{\rm EM}$ by the evection resonance.  Our results do not show a preference for creating final systems with an AM comparable to that of the Earth-Moon.  This contrasts with the findings of \cite{cuk2012making}, who argued that for a certain range of $A$, final systems with $L_{\rm F} \sim L_{\rm EM}$ would always result so long as $Q$ was large enough to allow initial capture into resonance. We instead find that the final system AM varies greatly even for a given $A$, for different $Q_{\rm eff}/k_{2\oplus}$ values. 

The Earth's mantle likely remained molten and fluid-like in its tidal response for few $\sim 10^6$ yr due to the Earth's thermally blanketing atmosphere, while the Moon would have more rapidly cooled to a partially solid, tidally dissipative state \citep{zahnle2015tethered}. Thus the physically most plausible conditions when the Moon encountered evection resonance are low $\Delta t$ (high $Q_{\rm eff}$) and high $A$ \citep{zahnle2015tethered}.  For this combination, and assuming fixed (i.e., non-time dependent) values for $\Delta t$ and $A$, we find that evection resonance and the subsequent QR-regime typically removes too much AM, and often leaves the system in the dual synchronous state in which the lunar month equals the terrestrial day.  The latter would ultimately lead to the loss of the Moon as the Earth's spin was further slowed by solar tides, causing the Moon to then lie inside co-rotation and be subject to inward tidal evolution.

However, tidal parameters may have instead changed with time during the Moon's interactions with evection  resonance and the subsequent QR-regime.  For low-$A$ cases that lead to large evection-resonance-driven lunar eccentricities, \cite{tian2017coupled} showed that tidal heating in the Moon causes rapid escape from evection resonance as the lunar tidal $Q$-value decreases, resulting in insufficient AM removal.

For the high-$A$ cases advocated here and in \cite{zahnle2015tethered}, excitation of the lunar eccentricity by evection resonance is modest, and tidal heating in the Moon would be unlikely to substantially affect the lunar tidal parameters \citep{tian2017coupled}.  For high-$A$ cases, it is instead the time evolution of the Earth's $Q$ value as its mantle began to freeze that is important to consider.  We adopt the time-dependent $Q$ model of \cite{zahnle2015tethered}, and find that escape from QR occurs early, so that the final system is left with a\add{n} AM substantially larger than in the Earth-Moon.  \cite{zahnle2015tethered} point out that the predicted $Q$  as a function of the Moon's semi-major axis (or alternatively, $Q$  as a function of time) is essentially constant across many assumed atmospheric and albedo conditions, so long as the terrestrial mantle after the giant impact was sufficiently molten that it responded like a low viscosity liquid.  Because the latter appears highly probable \citep{Nakajima2015286}, a successful outcome in which the final system AM is $\sim L_{\rm EM}$ after evection resonance and a subsequent QR-regime may be difficult to achieve.  It is, however, possible that a constant lag angle model could yield different results than those seen here with Mignard tides, as the effect of a time-varying terrestrial $Q$ was not considered in the limit cycle analysis of \cite{tian2017coupled}.

Additional AM removal processes (e.g., solar tides; \citealp{canup2004simulations}, late veneer impacts; \citealp{bottke2010stochastic}, core/mantle friction \citealp{goldreich1970obliquity}) appear to have a small effect on the final AM of the system. Alternatively, if Earth's obliquity was $>65^{\circ}$ after the Moon-forming impact, the Moon could encounter an instability during the Laplace plane transition (occurring when the effects of Earth's oblateness on the lunar precessional motion are comparable to those of the Sun). During the instability, Earth's spin decreases and AM is transferred to Earth's orbit \citep{Cuk2016}.  However, \cite{tian2020vertical} argue that such an evolution cannot reproduce the current Earth-Moon system, because the needed initial high-obliquity state is inconsistent with the component of the current Earth-Moon angular momentum that is perpendicular to the ecliptic plane.

\nolinenumbers
\section*{Acknowledgements}
This paper is dedicated to the memory of William R. Ward. This research was supported by NASA's SSERVI and Emerging Worlds programs. RR is an Awardee of the Weizmann Institute of Science - National Postdoctoral Award Program for Advancing Women in Science. We thank the anonymous reviewers for their comments that improved the final version of this manuscript. The equations provided in the main text and in the referenced works provide all information
needed to reproduce the results presented in the manuscript. \add{The numerical results presented in Figure 6 and 8 are provided in Tables S2 and S3, respectively.}

\renewcommand\refname{Supporting References}

\end{document}


\supportinginfo{Tidal Evolution of the Evection Resonance/Quasi-Resonance and the Angular Momentum of the Earth-Moon System}

\authors{R. Rufu  and R. M. Canup}
\affiliation{}{Planetary Science Directorate, Southwest Research Institute, Boulder, Colorado, 80302, USA}

\correspondingauthor{R. Rufu}{raluca@boulder.swri.edu}

\section*{Contents}
\begin{enumerate}
\item Text S1 to S2
\item Table S1
\item Caption Table S2 to S3
\item Figure S1 to S8

\end{enumerate}
\section*{Text S1 - Mignard Model}

For a given time delay in the Mignard tidal model, the resulting "lag angle" on the perturbed body (the angle between the tide and the line connecting the centers of the Earth and Moon) varies with frequency.  The corresponding orbit-averaged lag angle for the tide raised on the Earth by the Moon is $\delta_{\oplus} = (s-n) \Delta t$, where $s$ is the Earth's angular spin velocity and $n$ is the orbital mean motion.  Analogously, the orbit-averaged lag angle for the tide raised on the Moon by the Earth is $\delta_m = (s_m-n)\Delta t_m$, where $s_m$ is the Moon's angular spin velocity. For $s = n$, i.e., the Moon orbiting at the co-rotation radius, $\delta_\oplus = 0$.  Analogously, for $s_m = n$, corresponding to a lunar rotation that is synchronous with the Moon's mean motion, $\delta_m = 0$. A strength of the Mignard model is that its predicted tidal torques vary smoothly as the relevant frequencies change, producing physically realistic values even when $\delta_E = 0$ or $\delta_m = 0$.  This is a key advantage over constant lag angle models, which encounter discontinuities when passing through the $s = n$ or $s_m = n$ commensurabilities.

Consider the simplest case of a circular lunar orbit ($e = 0$) and we set $\mu+1\sim1$. Mignards' non-normalized expression for the orbit-averaged change in the Moon's semi-major axis due to Earth tides for this case reduces to:
\begin{equation}
    \left.\frac{da}{dt}\right\vert_{\oplus} = 6 k_{2\oplus}\mu\Delta t \Omega_\oplus R_\oplus (s-n) \left(\frac{a}{R_\oplus}\right)^{-11/2}
\end{equation}
so that for $s=n$, there is no torque on the lunar orbit due to Earth tides.  
Analogously, $da/dt$ of the lunar orbit due to Moon tides is:
\begin{equation}
     \left.\frac{da}{dt}\right\vert_{m} = 6 k_{2\oplus}\mu\Delta t \Omega_\oplus R_\oplus A(s_m-n)\left(\frac{a}{R_\oplus}\right)^{-11/2},
\end{equation} 
so that if $s_m = n$, there is no torque on the lunar orbit due to lunar tides for $e = 0$.  The $s_m = n$ case is commonly considered, because the time for a satellite's rotation to synchronize with its orbit around a much more massive primary is often short.  The lunar tide produced when $s_m = n$ and $e\neq0$ is often referred to as the radial push-pull tide.  Because in this case $\delta_m=0$, there is no torque on the lunar orbit due to the lunar tide. However, the variation in the magnitude of the tide from periapse to apopase changes the height of the tidal bulge during each orbit, with the maximum lunar tide occurring somewhat after the pericenter due to the time lag. This leads to energy dissipation within the satellite, which is accommodated by a change in its orbital energy at a constant orbital angular momentum (\textit{e.g.}, \citealp{peale1986orbital}). Thus, even with $s_m = n$ (a synchronously rotating Moon), there will be changes in the Moon's orbit due to lunar tides if its orbit is non-circular. A push-pull tide would analogously occur in the Earth for $s=n$ if the Moon's orbit is eccentric, although because the Moon-to-Earth mass ratio is so small this effect is not important.

We consider the Mignard expressions due to lunar tides for a non-circular orbit and $s_m = n$, where for simplicity we consider only terms to second order in $e$ (so that $f_1-f_2\sim-19/2e^2$ and $g_1-g_2\sim-7/2-113/2e^2$).  In this case, it is straightforward to show that \cite{ward2020analytical} eqn. (4.9 a,b) for $de/dt$ and $da/dt$ due to lunar tides in the un-normalized form and the definition of $A$ from their eqn. (4.10) give 
\begin{equation}
    \left.\frac{da}{dt}\right\vert_{m} = -57ae^2(n\Delta t_m)k_{2m}n\mu^{-1}\left(\frac{R_m}{a}\right)^5
    \label{eq:dadt}
\end{equation}
\begin{equation}
   \left.\frac{de}{dt}\right\vert_{m} = -\frac{21}{2}e(n\Delta t_m)k_{2m} n\mu^{-1}\left(\frac{R_m}{a}\right)^5
   \label{eq:dedt}
\end{equation}
The Mignard time delay can be related to a tidal quality factor, $Q$, as $Q \sim (\psi \Delta t)^{-1}$ for a system oscillating at frequency $\psi$ (e.g., \citealp{peale2015origin}).  For a synchronously rotating Moon, the appropriate frequency for lunar tides is $\psi=n$, so that $n\Delta t_m = 1/Q_m$, where $Q_m$ is the lunar tidal dissipation. The equations above are then equal to the standard forms in \cite{goldreich1966q} (their eqn. 25), \cite{peale1986orbital} (eqn. 47; see also discussion in  \citealp{peale2015origin}, Appendix B, second term in their eqn. B-28), and in \cite{Wisdom2015138} (their eqn. 18-19, terms that are dependent on $A$, as defined in their eqn. 12, and represent the satellite tides).  Thus to the second order in eccentricity, the Mignard formalism and its associated definition of $A$ (eqn. 9 in the main paper) predicts rates of change in $a$ and $e$ due to lunar tides that are the same as in the constant-$Q$ tidal model for the case of synchronous lunar rotation. 

However, for a non-circular orbit, one must also include the effects of torques on the Moon's figure necessary to maintain synchronous rotation. Tides have a net torque on the lunar spin that tend to move the Moon away from synchronous rotation for $e\ne0$. The torque on a tri-axial lunar shape can maintain a synchronous spin even when the orbit is substantially eccentric \citep{goldreich1966spin}. The figure torque leads to additional changes in the semimajor axis and eccentricity, which were not included in the original \cite{mignard1980evolution} model. 

Terms including the permanent figure torque, assuming it is sufficient to maintain synchronous rotation, are derived in \citeauthor{ward2020analytical} (\citeyear{ward2020analytical}, Appendix D).  Including only terms up to second order in $e$, the figure torque modifies $da/dt$ as (modifications to $de/dt$ are higher order in $e$) 
\begin{equation} 
    \left.\frac{da}{dt}\right\vert_{ pf} = 36ae^2(n\Delta t_m)k_{2m} n\mu^{-1}\left(\frac{R_m}{a}\right)^5
\end{equation}

The expression above is the same (to the second order eccentricity) as the equivalent permanent figure torque corrections in \citeauthor{williams2001lunar} (\citeyear{williams2001lunar}, their eqn. 35-36) and \citeauthor{Wisdom2015138} (\citeyear{Wisdom2015138}, their eqn. 21-22). 

\section*{Text S2 - Numerical Tests}
To determine the sensitivity of the minimum eccentricity limit in our code, we performed a series of simulations with varied minimum eccentricities, a constant relative tidal factor $A=500$, and $Q_{\rm eff}/k_{2\oplus}\sim410$. We used a high $A$ in order to insure that the eccentricity derivative initially is negative and the minimum eccentricity is reached before entering the evection phase.
For low minimum eccentricities ($<1e-35$), or alternatively when the  eccentricity is not limited, the system passes evection, but the AM remains high, as eccentricities are numerically not resolved. However, for a large range of eccentricities ($10^{-35}$ to $10^{-4}$), the final AM is not sensitive (variations $<3\%$ in final AM) to the minimum eccentricity value chosen. We implemented a minimum eccentricity of $10^{-8}$, which is consistent with expected excitation due to interactions with the leftover background planetesimal population (e.g., \citealp{spurzem2009dynamics, Pahlevan:2015aa}).

During evection proper, the amplitude of oscillations around the stable stationary eccentricity grows when the orbit begins to contract, and eventually causes the system to escape the resonance. Resolving these oscillations is critical to determining the timing of resonance escape. The libration frequency is linearly proportional to the tidal timescale and eccentricity ($\omega\propto t_Te_s$; \citealp{ward2020analytical}), hence the time step of the integration (or relative tolerance for adaptive-step techniques) must be adjusted as a function of the absolute tidal rate assumed, such that the librations at the maximum eccentricity are resolved. 

\begin{figure} 
    \centering
    \includegraphics[width=0.8\linewidth]{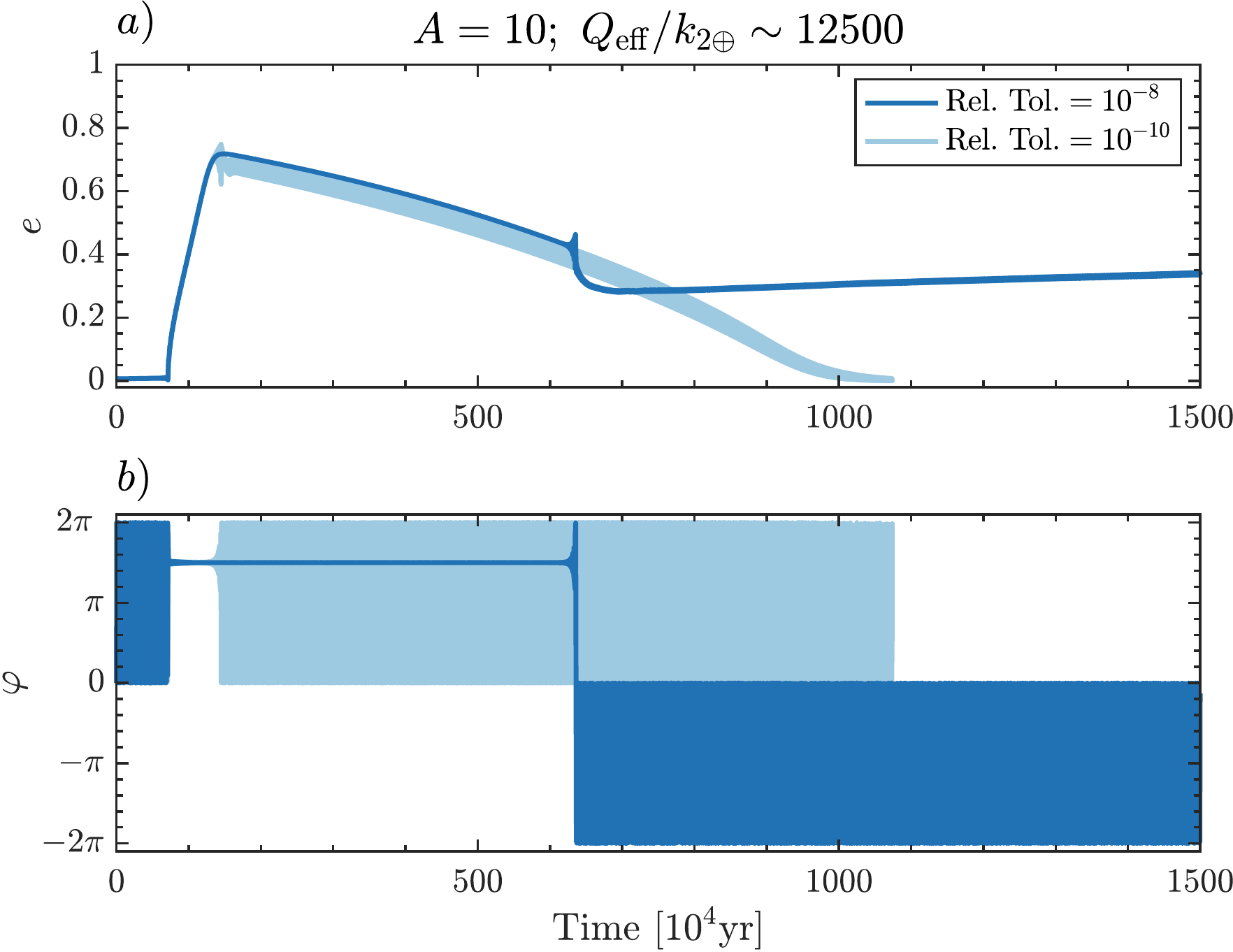}
    \caption{\raggedright Tidal evolution of the Earth-Moon system for $A=10$, terrestrial tidal parameters $Q_{\rm eff}/k_{2\oplus}\sim 1e5$, with different relative tolerances for setting the integration time step ($\rm Rel.\ Tol.$). a) Eccentricity; b) Resonance angle as a function of time. For the higher relative tolerance (dark blue), the librations around the stationary point are not resolved, and the system appears to remain in resonance (constant angle) for a protracted period after the orbit contracts. This is similar to the outcome seen in \cite{cuk2012making} (although they used a different tidal model). With a lower $\rm Rel.\ Tol.$ the librations are adequately resolved and they cause escape from resonance proper near the time of orbit contraction, consistent with analytic predictions of \cite{ward2020analytical}.}
     \label{fig:LateRes}
\end{figure}

If the time step is not adequately small, the system can erroneously remain in resonance after the orbital contraction (dark blue in Fig. \ref{fig:LateRes}, similar to the behaviour reported by \citealp{cuk2012making}, and in preliminary results in \citealp{rufu2019evection}). As the orbit further contracts, eccentricity decreases, until the oscillation timescale is comparable to the integration time step. Instantly after that, the libration amplitude grows and exit from evection proper occurs. 

On the other hand, if the time step is sufficiently small, oscillations during evection proper are resolved, and once the orbit contracts, the libration amplitude grows and exit from resonance proper occurs. Although \cite{ward2020analytical} used a simplified model for estimating when libration amplitude growth would occur, the prediction of the timing of evection escape in that work is in agreement with our numerical simulations.

\begin{figure} 
\centering
    \includegraphics[width=0.8\linewidth]{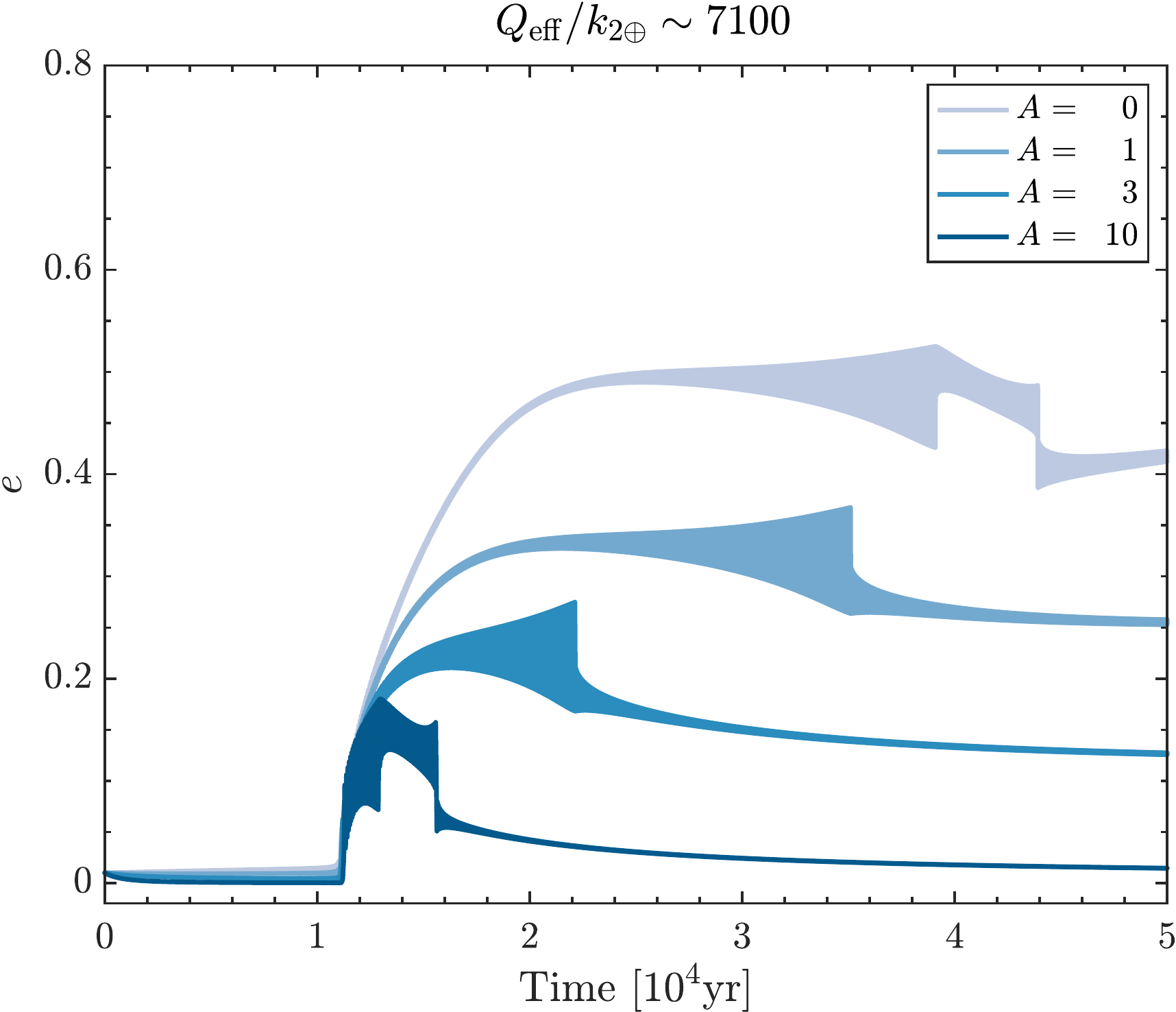}
    \caption{\raggedright Eccentricity evolution with different relative tidal strength values, $A$, and assuming  terrestrial tidal parameters $Q_{\rm eff}/k_{2\oplus}\sim 7100$. Similarly to \cite{touma1998resonances}, for this set of simulations we assumed an initial terrestrial spin of $5$ hr and a synchronous lunar spin (the additional terms required to maintain synchronicity for high-$e$ orbits were not included). The eccentricity evolution calculated using the semi-analytical model developed in \cite{ward2020analytical} and used in this work produces comparable results to the N-body calculations of \citeauthor{touma1998resonances} (\citeyear{touma1998resonances}, Fig. 7 in their paper). The timing disparities result from a different chosen $\Delta t$. }
    \label{fig:TWcomp}
\end{figure}

We have compared results of our numerical model to those of previous simulations by \cite{touma1998resonances}, which used the Mignard model \citep{mignard1980evolution}. As in \cite{touma1998resonances}, for this set of simulations we assumed an initial terrestrial spin of $5$ hr, a synchronous lunar rotation, and we did not include terms associated with permanent figure torques. The eccentricity evolution calculated using our semi-analytical model (Fig. \ref{fig:TWcomp}) is similar to that obtained from N-body calculations \citep{touma1998resonances}. The timing disparities between the two studies result from different $\Delta t$  values, which control the lunar tidal acceleration.

\newpage
\setlength{\arrayrulewidth}{1pt}
\begin{table}[htb]
     \centering 
      \caption{Tidal polynomial functions\label{PolFun}}
\vspace{-3mm}

\begin{tabular}{l}
\hline
    $\tilde{f}_1(e) = 1+15/2e^2+45/8e^4+5/16e^6    $   \\
    $\tilde{f}_2(e) = 1+31/2e^2+255/8e^4+185/16e^6+25/64e^8$   \\
    $\tilde{g}_1(e) = 11/2+33/4e^2+11/16e^4$    \\
    $\tilde{g}_2(e) = 9+135/4e^2+135/8e^4+45/64e^6$\\
    \\
    $f_1(e) = \tilde{f}_1(e)/(1-e^2)^6$ \\
    $f_2(e) = \tilde{f}_2(e)/(1-e^2)^{15/2}$ \\
    $g_1(e) = \tilde{g}_1(e)/(1-e^2)^5$ \\
    $g_2(e) = \tilde{g}_2(e)/(1-e^2)^{13/2}$ \\
\end{tabular}
\end{table}

\section*{Caption Table S2}
Final angular momentum values after the lunar passage through evection resonance and the subsequent quasi-resonance (QR) for the non-synchronous lunar spin cases. $Q_{\rm eff}/k_{2\oplus}$, Earth's tidal dissipation parameter (column \#1); $A$, the relative tidal strength factor (column \#2); $L^{\prime}_{\rm{F}_{\rm QR}}$, the normalized minimum final AM in each combination, obtained in the simulations that exited on the QR side (column \#3, also described in Fig. 6); $L^{\prime}_{\rm{F}_{\rm Non-QR}}$, the normalized minimum final AM in each combination, obtained in the simulations that exited on the non-QR side (column \#4); $\rm{Frac}_{\rm{QR}}$, the fraction of simulations that exited the resonance on the QR side (column \#5). NaN values in column \#3 [\#4] depict cases where a QR [non-QR] evolution did not occur.
\section*{Caption Table S3}
Similar to Table S2 for cases that include the permanent figure torque (synchronous lunar spin; depicted in Figure 8).


\newpage

\begin{figure} 
    \centering
    \includegraphics[width=0.9\linewidth]{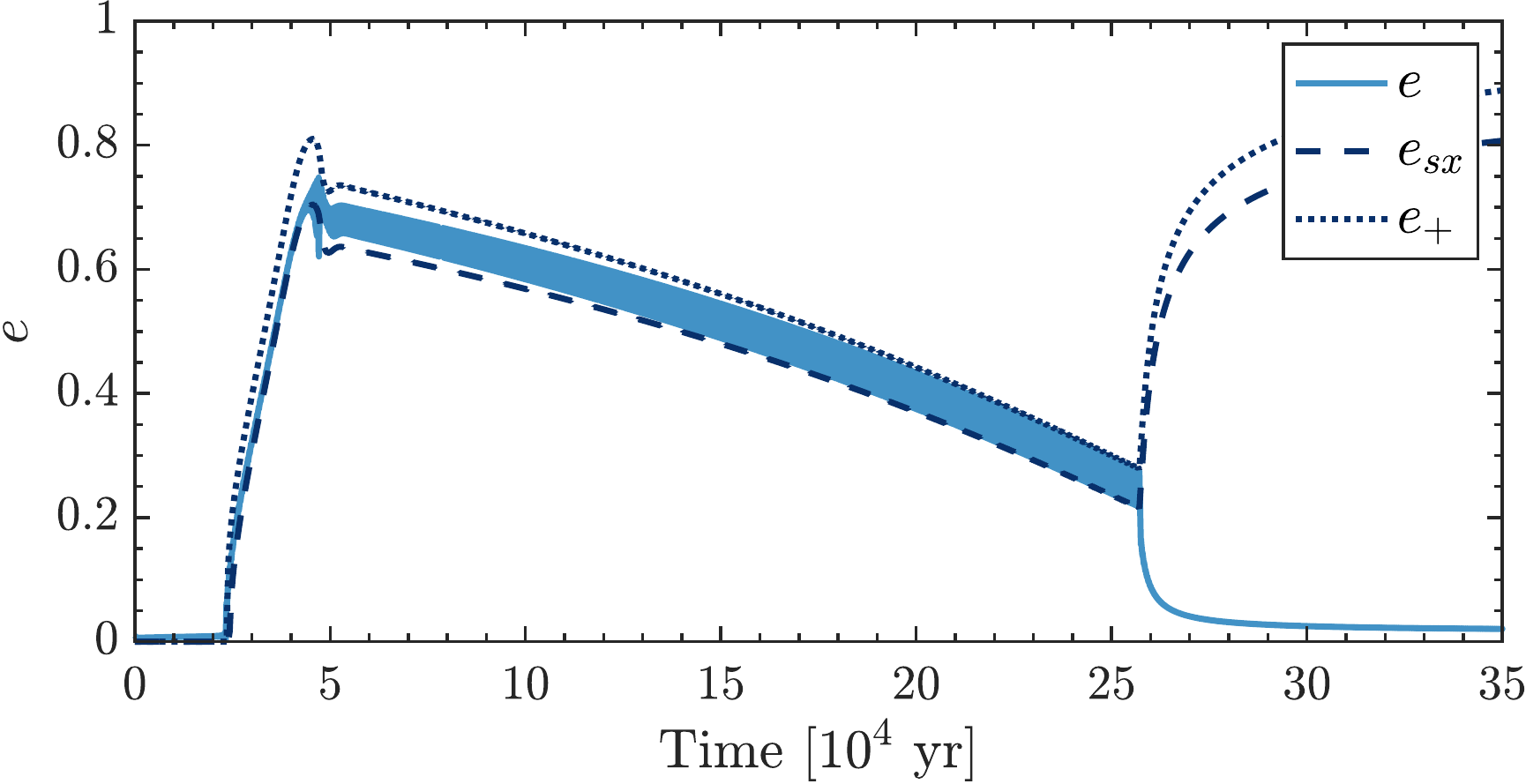}
    \caption{\raggedright Lunar eccentricity (solid line) evolution during evection and QR for the evolution shown in Fig. 3 ($A=10$ and $Q_{\rm eff}/k_{2\oplus}\approx 410$). During QR, the eccentricity oscillates between a minimum value close to the unstable stationary eccentricity ($e_{sx}$; dashed line) and a maximum value comparable to that of the outer separatrix curve in the phase diagram ($e_+$; dotted line).}
     \label{fig:EccEsxE+}
\end{figure}
\begin{figure}
    \centering
    \includegraphics[width=0.8\linewidth]{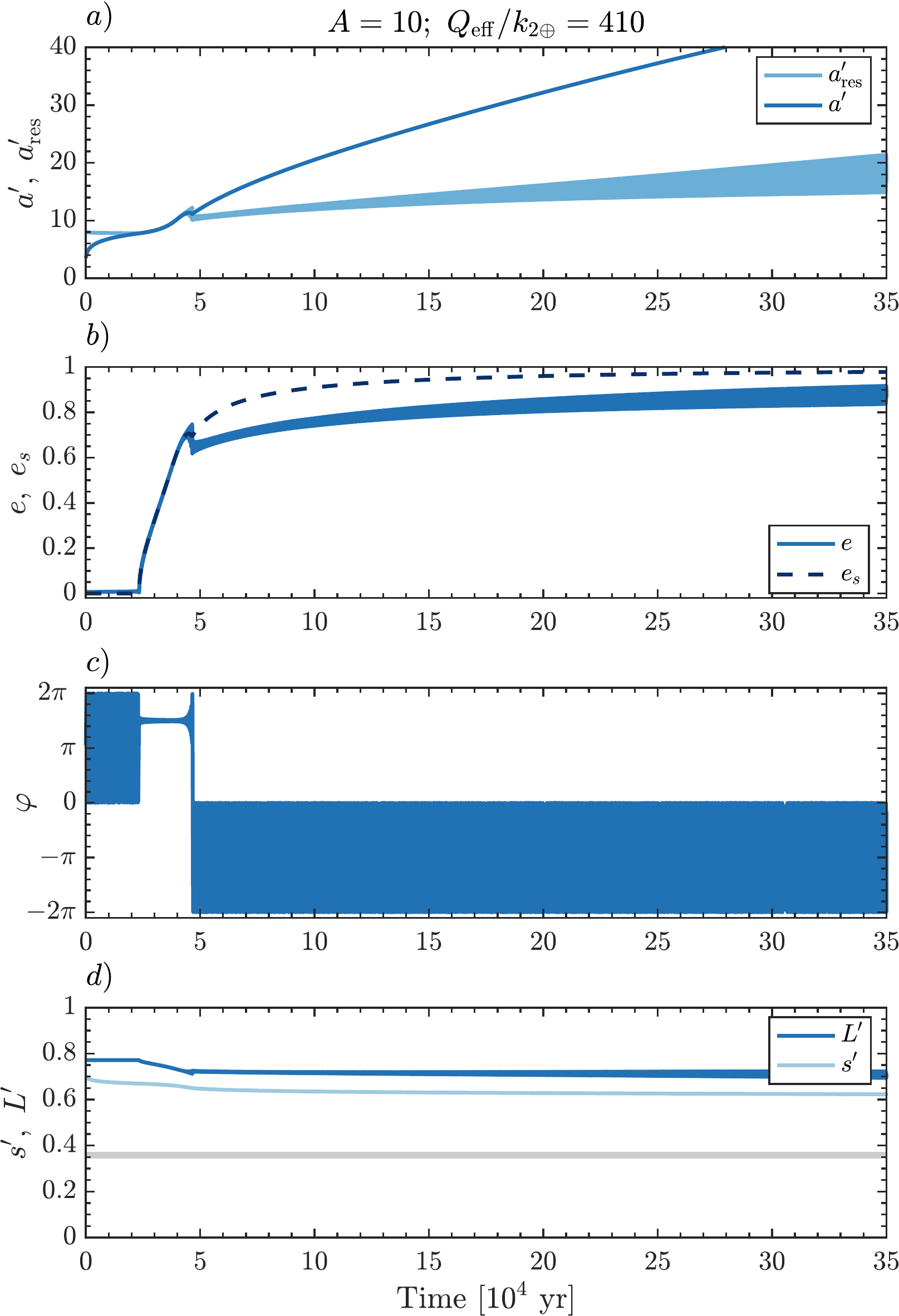}
    \caption{\raggedright Same as Fig. 3 with a different $\varphi(0)$. The exit from resonance occurs below the stable stationary eccentricity, so that there is no QR phase and only minimal AM is removed.}
    \label{figSup:EvolutionSlowTidesPhi5}
\end{figure}

\begin{figure} 

    \centering
    \includegraphics[width=0.8\linewidth]{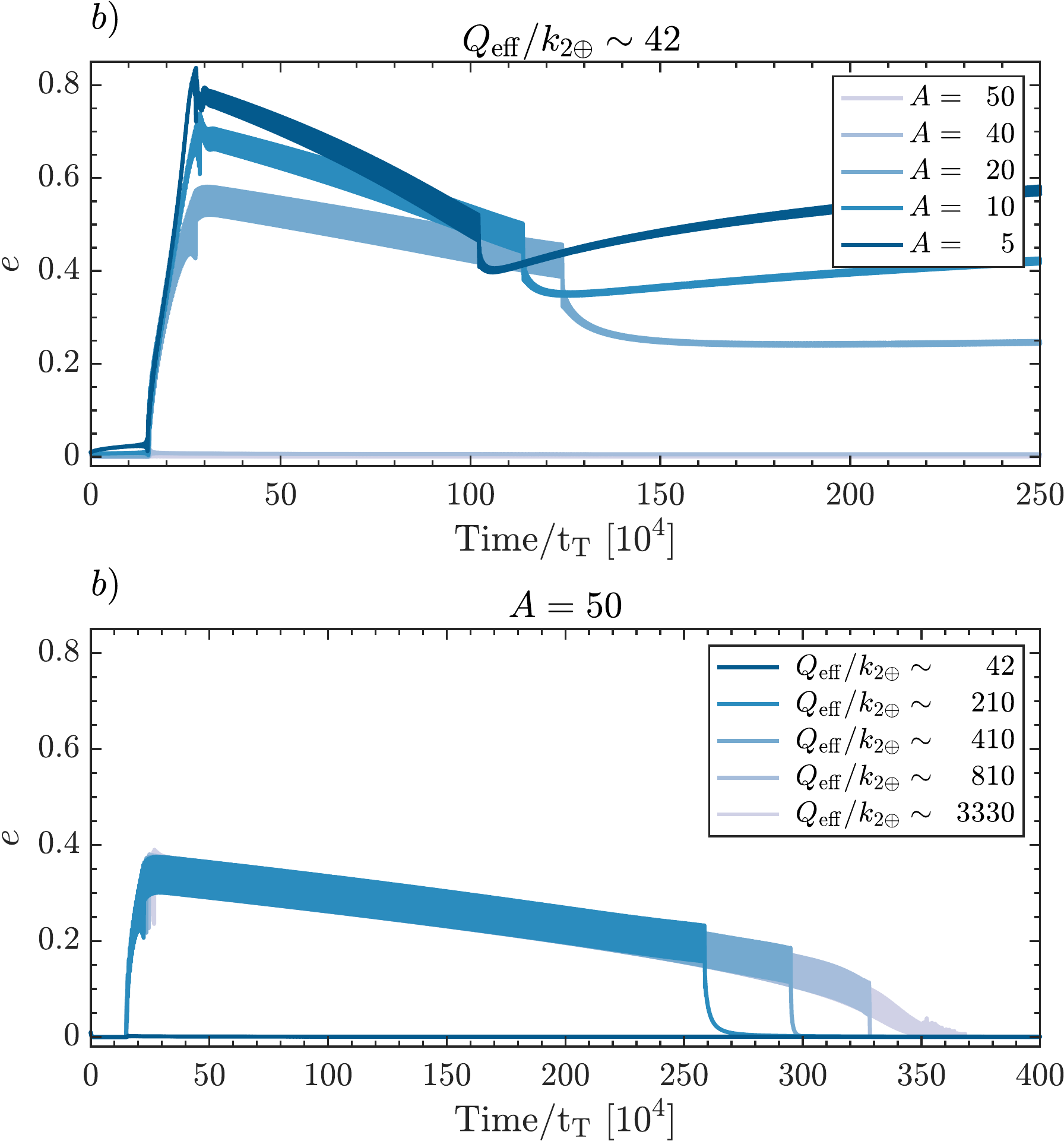}

    \caption{\raggedright Eccentricity evolution with different a) relative tidal strength values, $A$ (assuming a small terrestrial tidal parameter  $Q_{\rm eff}/k_{2\oplus}\sim42$, or equivalently a fast terrestrial tidal rate); b) terrestrial tidal parameter, $Q_{\rm eff}/k_{2\oplus}$ (assuming $A=50$). The time is normalized by the tidal timescale, $t_T$.  For $A\geq40$ or $Q_{\rm eff}/k_{2\oplus}\leq42$ the system exits proper evection when $e\sim10^{-3}$, before the resonance is fully developed. This occurs because the upward movement of the stable stationary point due to tides is faster than the libration around that point (see section 3.2 for details). }
     \label{fig:DiffA}
\end{figure}

\begin{figure} 

    \centering
    \includegraphics[width=0.8\linewidth]{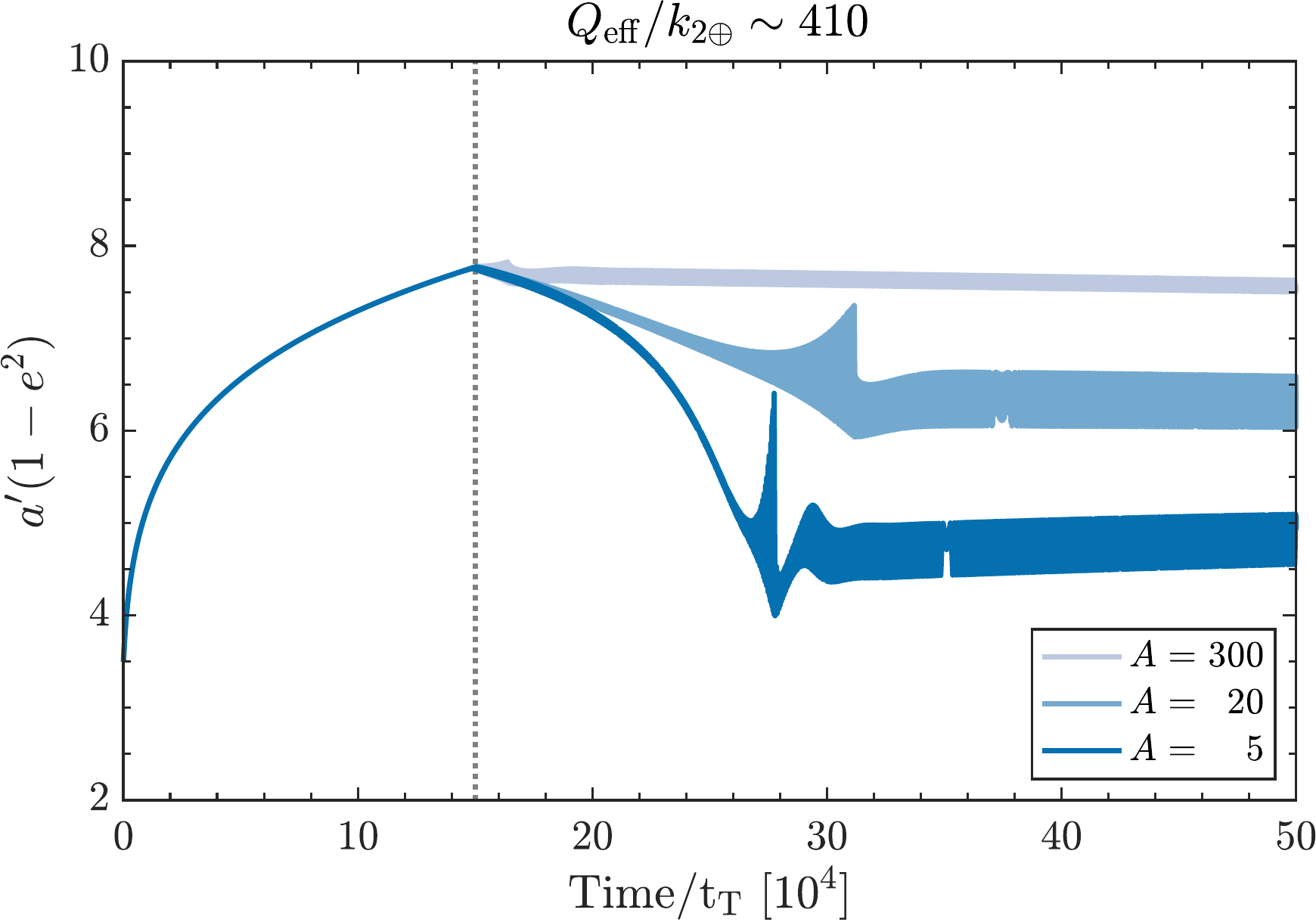}

    \caption{\raggedright The evolution of $a^{\prime}(1-e^2)$ with different relative tidal strength values, $A$ (assuming a small terrestrial tidal parameter  $Q_{\rm eff}/k_{2\oplus}\sim410$). The time is normalized by the tidal timescale, $t_T$. The initial capture into evection occurs at $15\times10^4\,t_T$ (dashed line). Exit from evection proper varies across different $A$ values (e.g. for $A=5$, exit occurs at $\sim26\times10^4\,t_T$). With lower  $A$ values, $a^{\prime}(1-e^2)$  at the time of escape from proper resonance decreases, therefore the total AM at the end of proper resonance decreases ($L^{\prime}\approx  a^{\prime7/4}(1-e^2)/\Lambda+\gamma \sqrt{a^{\prime}(1-e^2)}$).}
     \label{fig:DiffA}
\end{figure}  


\begin{figure} 

    \centering
    \includegraphics[width=1\linewidth]{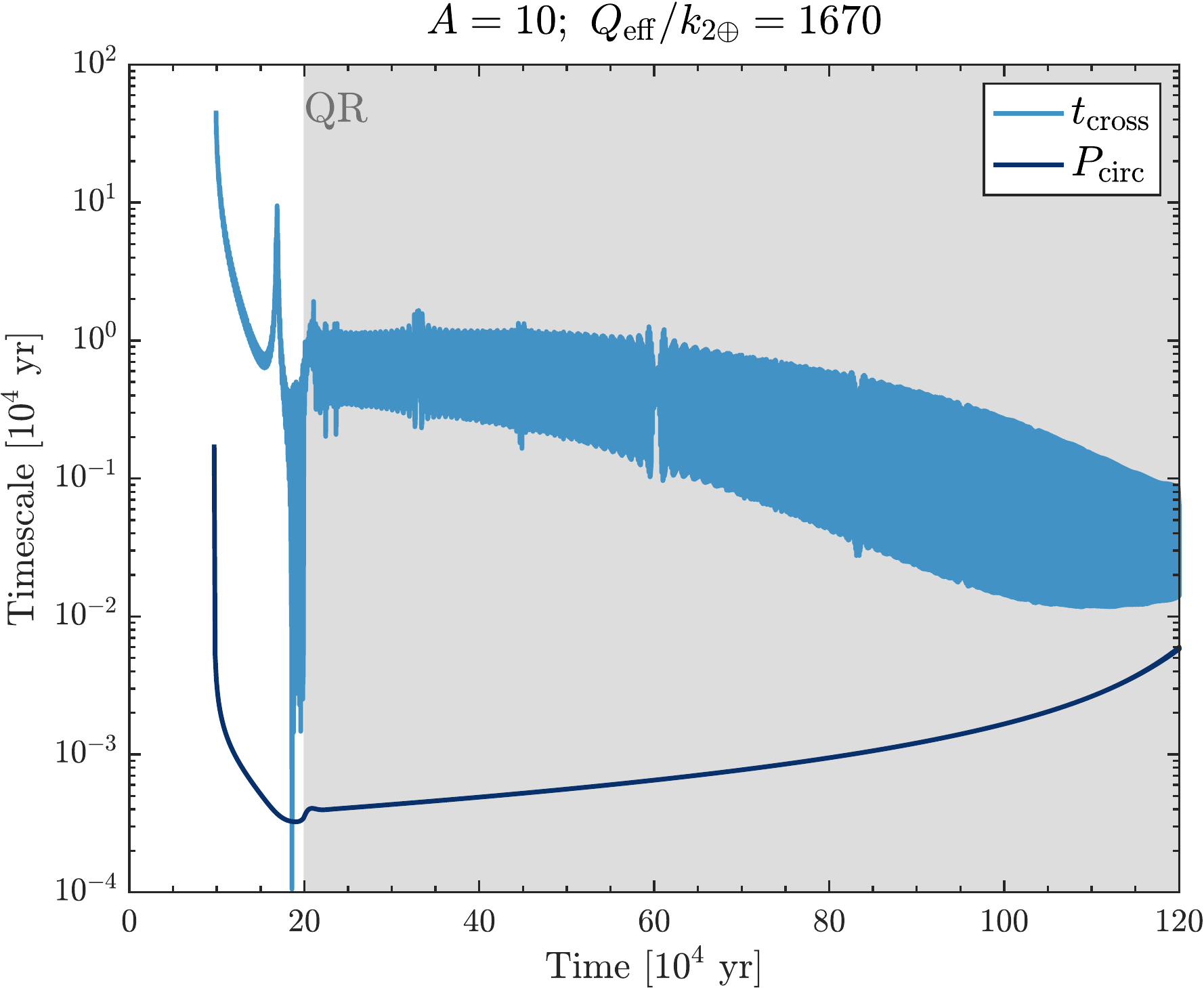}

    \caption{\raggedright The timescale required for tides to push the eccentricity across the separatrix ($t_{\rm cross}$; light blue) compared to the timescale of circulation ($P_{\rm circ}$; dark blue) for an evolution that remained in QR until the co-synchronous state ($A=10$ and $Q_{\rm eff}/k_{2\oplus}=1670$). The grey area represents the time the system is in the QR phase. At $\sim20\cdot10^4\ {\rm yr}$, the libration amplitude increased and the system is pushed from the librating resonant area across the separatrix, exiting proper evection. The system subsequently remains in QR until the dual synchronous state is reached. Similar outcomes would be expected for larger $Q_{\rm eff}/k_{2\oplus}$ values too.}
     \label{fig:Sup_CrossLib}
\end{figure}  

\begin{figure} 

    \centering
    \includegraphics[width=1\linewidth]{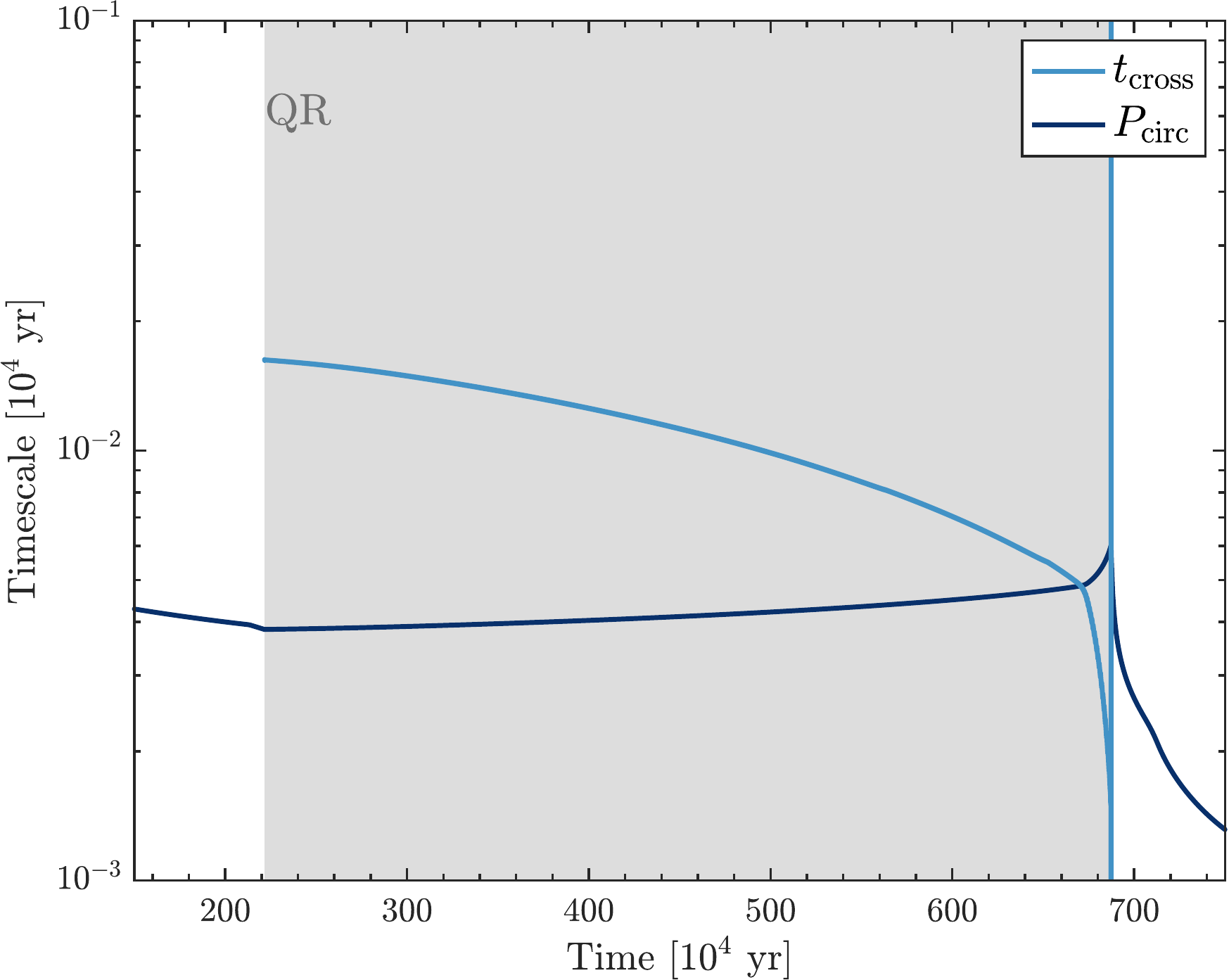}
    \caption{\raggedright The timescale required for tides to drive the eccentricity across the separatrix ($t_{\rm cross}$, light blue) compared to the timescale of circulation ($P_{\rm circ}$; dark blue) for the evolution depicted in Fig. 9-b during Earth's cooling. The grey area represents the time the system is in QR phase. When the two timescales are comparable, the trajectory is pushed inwards across the separatrix in a time comparable to the circulation time.}
     \label{fig:CrossLibVariableTau}
\end{figure}